\documentclass[preprint,showpacs,preprintnumbers,amsmath,amssymb]{revtex4}

\usepackage{epsfig}
\usepackage{graphics}
\usepackage{graphicx}
\usepackage{dcolumn}
\usepackage{bm}

\begin{document}

\title{Delay time modulation induced oscillating synchronization and 
intermittent anticipatory/lag and complete synchronizations in 
time-delay nonlinear dynamical systems }

\author{D.~V.~Senthilkumar}
\email{skumar@cnld.bdu.ac.in}
\author{M.~Lakshmanan}%
 \email{lakshman@cnld.bdu.ac.in}
\affiliation{%
Centre for Nonlinear Dynamics,Department of Physics,
Bharathidasan University, Tiruchirapalli - 620 024, India\\
}%

\date{\today}

\begin{abstract}  
Existence of a new type of oscillating synchronization that oscillates between
three different types of synchronizations (anticipatory, complete and lag
synchronizations) is identified in  unidirectionally coupled nonlinear
time-delay systems  having two different time-delays, that is feedback delay 
with a periodic delay time modulation and a constant coupling delay.
Intermittent anticipatory, intermittent lag and complete synchronizations are
shown to exist in  the same system with identical delay time modulations in
both the  delays. The transition from anticipatory to complete synchronization
and from complete to lag synchronization as a function of coupling delay with
suitable stability condition is discussed. The intermittent anticipatory and
lag synchronizations are characterized by the minimum of similarity functions
and the intermittent behavior is characterized by a universal asymptotic
$-\frac{3}{2}$ power law distribution.  It is also shown that the delay time
carved out of the trajectories of the time-delay system with periodic delay
time modulation cannot be estimated using conventional methods, thereby
reducing the possibility of decoding the message by phase space
reconstruction. 
\end{abstract}

\maketitle
{\bfseries Synchronization of chaos is one of the most fundamental phenomena
exhibited by coupled chaotic oscillators. Recent studies on chaotic
synchronization has also focussed on nonlinear time-delay systems in view of
their hyperchaotic nature.  Further, the concept of delay time modulation has
been introduced in order to understand dynamical systems with time
dependent topology such as internet, world wide web, population dynamics,
neurology, etc. It has also been shown that  nonlinear delay systems with time
dependent delay can exhibit more complex dynamics. Consequently, studies on
synchronization of such systems with time dependent delay becomes very
important in order to understand their cooperative dynamics. From this point of
view, in this paper we have considered simple scalar piecewise linear
time-delay systems with unidirectional time-delay coupling in order to explore
the various types of synchronized behaviours and their transitions. The
introduction of  simple fully rectified sinusoidal modulation in these systems with constant
coupling delay can lead to the existence of a new type of oscillating
synchronization that oscillates between three different types of
synchronizations, namely complete, lag and anticipatory synchronizations. With
delay time modulation in both the coupling and feedback delay, the coupled
system displays intermittent lag/anticipatory and complete synchronizations
for  suitable ranges of values of the delay times and modulational frequencies.
It has also been shown that  the decrease in the value of modulational
frequency leads to exact and then to lag/anticipatory synchronizations from
their intermittent nature. The existence of different synchronizations are
corroborated by  suitable stability condition based on Krasovskii-Lyapunov
theory and their corresponding similarity functions.  The intermittent regimes
are characterized by a  universal asymptotic $-\frac{3}{2}$ power law
distribution.}

\section{\label{sec:level1}Introduction}
Chaos synchronization has been receiving a great deal of interest  for more
than two decades in view of its potential applications in various fields of
science
\cite{hfty1983,hfty198370,lmptlc1990,eocg1990,aspmgr2001,mlkm1996,mlsr2003}. 
Since the identification of chaotic synchronization,  different kinds of
synchronizations have been proposed in interacting chaotic systems, which have
all been identified  both theoretically and experimentally.  Complete
synchronization refers to a perfect locking of chaotic trajectories so as to
remain in step with each other in the course of time,
$X(t)=Y(t)$~\cite{hfty1983,hfty198370,lmptlc1990,eocg1990}.  Generalized
synchronization is defined as the presence of some functional relationship
between the states of the drive and the response,
$Y(t)=F(X(t))$~\cite{nfrmms1995,rb1998,lkup1996}. Phase synchronization is
characterized by entrainment of  the phases of the two signals,
 whereas their
amplitudes remain uncorrelated~\cite{mgrasp1996,tyycl1997}.  Lag
synchronization implies that there is an exact time shift between the evolution
of drive and response systems, where the response lags the state of the drive,
$Y(t)=X(t-\tau)$~\cite{mgrasp1997,srik2002,mzgww2002}.  Anticipating
synchronization also appears as a coincidence of shifted-in-time states of the
two systems, where the response anticipates the state of drive,
$Y(t)=X(t+\tau)$~\cite{huv2002,huv2001,cm2001,ssems2001,emsss2002}.  Recently
intermittent lag synchronization has also been
identified~\cite{sbdlv2000,dlvsb2001,stycl1999}. Also it has been shown very
recently in a three-element network module of semiconductor laser system
that dynamical relaying can lead to zero-lag synchronization even in the
presence of coupling delays ~\cite{ifrv2006}.

The notion of time dependent delay (TDD) with stochastic or chaotic modulation
in time-delay systems was introduced by Kye \emph{et al} \cite{whkmc2004pre} 
to understand the behaviour of dynamical systems with time dependent topology. 
These authors have reported that in a time-delay system with TDD, the
reconstructed phase trajectory does not collapse to a simple manifold, a
property different from that of delayed systems with fixed delay time (which is
considered to be a serious drawback of the later type of systems).  It has been
shown very recently that a distributed delay enriches the characteristic
features of the delayed system over that of the fixed delay systems
\cite{fma2003}.  Based on these considerations, current studies on chaotic
synchronization in time-delay systems are focused towards time-delay systems
with time dependent delay
\cite{whkmc2004,whkmc2005pre,emskas2005,nlinemskas2005}.  In this connection it
is also of considerable interest to study the effect of simple modulations such
as periodic modulation  \cite{emskas2005,nlinemskas2005} on the nature of the
chaotic attractor.

Recently, we have studied chaotic synchronization in a system of two
unidirectionally coupled  odd piecewise linear time-delay
systems~\cite{dvskml2005} with two different constant delay times: one in the
coupling term and the other in the individual systems, namely, feedback delay.
We have shown that there is a transition from anticipatory to lag
synchronization through complete synchronization as a function of a system
parameter with suitable stability criterion.   The present work was motivated
by the fact that whether there arises any new phenomenon due to the
introduction of periodic delay time modulation in the coupled time-delay system
we have studied earlier and  its effects on the various synchronization
scenario. Interestingly, we have found that even with simple periodic
modulation, the time-delay system cannot be collapsed into a simple manifold
and that the delay time cannot be extracted using standard methods. More
interestingly, we have also found that the fully rectified sinusoidal modulation of delay time
introduces a new type of oscillating synchronization that oscillates between
anticipatory, complete and lag synchronizations for the case of constant
coupling delay.  This is further  corroborated by suitable stability condition based on
Krasovskii-Lyapunov theory. Intermittent anticipatory and lag synchronizations 
are  also found to exist in the present system for the case of identical
modulation in both the coupling and feedback delay, for a range of
modulational  frequencies. In addition, we also find that there exist regions
of exact anticipatory and lag synchronizations for lower values of
modulational  frequencies.   The results have been corroborated by the nature
of similarity functions and the intermittent behavior by the probability
distribution of the laminar phase, satisfying universal $-\frac{3}{2}$ power
law behavior of on-off intermittency~\cite{npsmh1994,jfhnp1994}.

The plan of  the paper is as follows.  In Sec.~II, we introduce the scalar
piece-wise linear time-delay system with delay time modulation and explore the
dynamical change in the time series of the time-delay system due to delay time
modulation. In Sec.~III we have introduced a unidirectional time-delay coupling
with delay time modulation  between the two scalar time-delay systems and we
have identified the condition for stability of the synchronized state following
Krasovskii-Lyapunov theory.  A new type of oscillating synchronization that
oscillates between anticipatory, complete and lag synchronizations and vice
versa is shown to exist in Sec.~IV for the case of constant coupling delay and
with modulated feedback delay. In Sec.~V, we  have pointed  out the existence
of intermittent anticipatory synchronization when the strength of the coupling
delay is less than that of the feedback delay with identical modulations, while in
Sec.~VI, complete synchronization is realized when the two delays are equal. 
Intermittent lag synchronization is shown to set in when the coupling delay
exceeds the feedback delay in Sec.~VII.  In Sec.~VIII we very briefly indicate
the possibility of more complicated oscillating synchronizations in the case of
nonidentical modulations. Finally in Sec.~IX, we summarize our results.
\section{Piecewise linear time-delay system with delay time modulation and 
dynamical changes}

At first, we will introduce the single scalar time-delay system with piecewise
linearity in the presence of delay time modulation, which has been studied
in detail for its chaotic dynamics in references 
\cite{ptkm1998,dvskml2005ijbc,hlzh1996} with constant
time-delay.  Then some measures to estimate the delay time will be discussed both
in the presence and in the absence of delay time modulation to show that the
imprints of the delay time carved out of the time series of the chaotic
attractor are completely wiped out by the modulation of delay time.
\subsection{The scalar delay system}
We consider the following first order delay differential equation introduced
by Lu and He~\cite{hlzh1996} and discussed in detail by Thangavel \emph{et al}.
~\cite{ptkm1998},
\begin{eqnarray}
\dot{x}(t)&=&-ax(t)+bf(x(t-\tau)),
\label{eqonea}
\end{eqnarray}
where $a$ and $b$ are parameters, $\tau$ is the constant time-delay and 
$f$ is an odd piecewise linear function defined as
\begin{eqnarray}
f(x)=
\left\{
\begin{array}{cc}
0,&  x \leq -4/3  \\
            -1.5x-2,&  -4/3 < x \leq -0.8 \\
            x,&    -0.8 < x \leq 0.8 \\              
            -1.5x+2,&   0.8 < x \leq 4/3 \\
            0,&  x > 4/3 \\ 
         \end{array} \right.
\label{eqoneb}
\end{eqnarray}
Recently, we have reported \cite{dvskml2005ijbc} that system
(\ref{eqonea}) exhibits hyperchaotic behavior for the parameter values
$a=1.0, b=1.2$ and $\tau=25.0$ and the hyperchaotic nature was confirmed by
the existence of multiple positive Lyapunov exponents (see Figures 1 and 2 in
ref.~\cite{dvskml2005ijbc}).

Now, we wish to replace the time-delay parameter $\tau$ as a function of time 
for our present study, instead of the constant time-delay, in the form
\cite{emskas2005,nlinemskas2005} 
\begin{eqnarray}
\tau(t)&=&\tau_0+\tau_a\left|\sin(\omega t)\right|,
\label{eqonec}
\end{eqnarray}
where $\tau_0$ is the zero frequency component, $\tau_a$ is the amplitude and 
$\omega/\pi$ is the frequency of the modulation. Note that in the delay term,
we have introduced the fully rectified sinusoidal modulational form (absolute
of the sine term) so as to keep the delay time positive even for values of
$\tau_a>\tau_0$ so as to avoid acausality problem in Eq.~(\ref{eqonea}) for
negative values of $\tau$ when $\tau_a>\tau_0$. However, for values
of $\tau_0$ sufficiently greater than $\tau_a$ the rectification in the
modulation (\ref{eqonec}) is not required.

\subsection{Estimation of the effect of delay time modulation}

Recently, the concept of time dependent delay with stochastic or chaotic
modulation was introduced by Kye \emph{et al.} \cite{whkmc2004pre} in
the time-delay systems and they have shown in the case of Mackey-Glass system
that the delay time carved out of time series of the time-delay system is
undetectable by the conventional measures and hence any reconstruction of phase
space of the delayed system is hardly possible. This fact has motivated some
authors \cite{whkmc2004,whkmc2005pre,emskas2005,nlinemskas2005} to look for  delay  systems
with delay time modulation  as an ideal candidate for secure communication.

Interestingly we find here that even with a fully rectified sinusoidal delay
time modulation of the form (\ref{eqonec}), system (\ref{eqonea}) exhibits the
properties studied by Kye \emph{et al} with stochastic or chaotic modulation.
In order to demonstrate the effect of fully rectified sinusoidal delay time
modulation of the form (\ref{eqonec}) on the time series of the piecewise
linear time-delay system which we have considered here, we will calculate 
\emph{(1) filling factor} \cite{mjbthm1997}, \emph{(2) length of polygon line}
\cite{mjbmp1996} and \emph{(3) average mutual information}
\cite{whkmc2004pre,amfhls1986,hdia1996} both in the presence and in the absence
of delay time modulation and show how periodic modulation removes any imprints
of the time-delay.

\subsubsection{Filling factor}
\begin{figure}[ht]
\centering
\includegraphics[width=0.7\columnwidth]{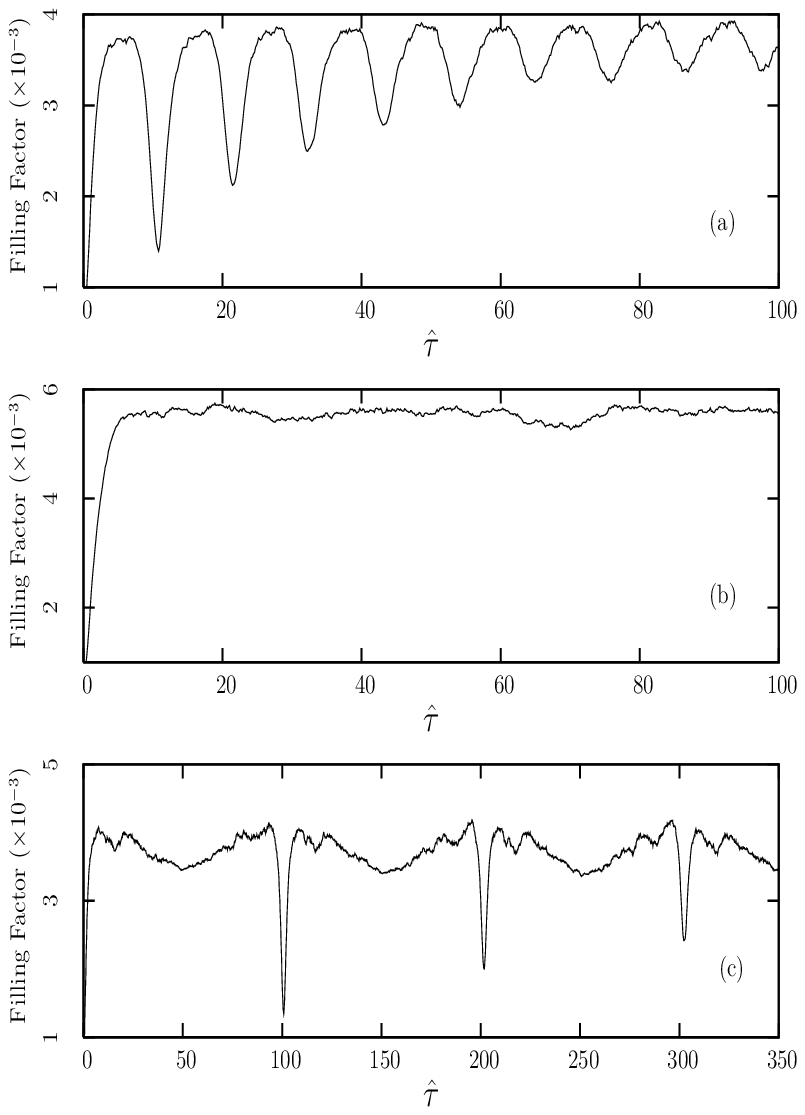}
\caption{\label{ff}}
\end{figure}

Now we will compute the filling factor \cite{mjbthm1997}  for the chaotic
trajectory $x(t)$ of the time-delay system (\ref{eqonea}) by projecting it onto
the pseudospace $(x,x_{\hat{\tau}},\dot{x})$ with $P^{3N}$ equally sized
hypercubes, where the delayed time series $x_{\hat{\tau}}=x(t-\hat{\tau})$ is
constructed from $x(t)$ for various values of $\hat{\tau}$. The filling factor
is the number of hypercubes which are visited by the projected trajectory,
normalized to the total number of hypercubes, $P^{3N}$. Figure~\ref{ff}a shows
the filling factor for constant delay $\tau_0=10$ when  $\tau_a=0$ in
Eq.~(\ref{eqonec}), where one can identify the existence of an underlying
time-delay induced instability \cite{mjbthm1997} which induces local minima in
the filling factor at $\hat{\tau}\approx n\tau_0,~ n=1,2,3...$. From the later,
one can identify the value of the time-delay parameter $\tau$ of the system
(\ref{eqonea}) under consideration. Figure~\ref{ff}b  shows filling factor with
delay time modulation of the form (\ref{eqonec}) with $\tau_0=10, \tau_a=90$
and $\omega=0.0001$, where no local minima occurs.  Figure~\ref{ff}c is plotted
for $\tau_0=100$ and $\tau_a=0$ to show that the disappearance of local minima
in Fig.~\ref{ff}b is not due to large delays but only because of delay time
modulation. From the figures one can realize that the imprints of the delay
time embedded in the projected trajectory is completely ironed out due to the
presence of delay time  modulation.

\subsubsection{Length of polygon line}
\begin{figure}
\centering
\includegraphics[width=0.7\columnwidth]{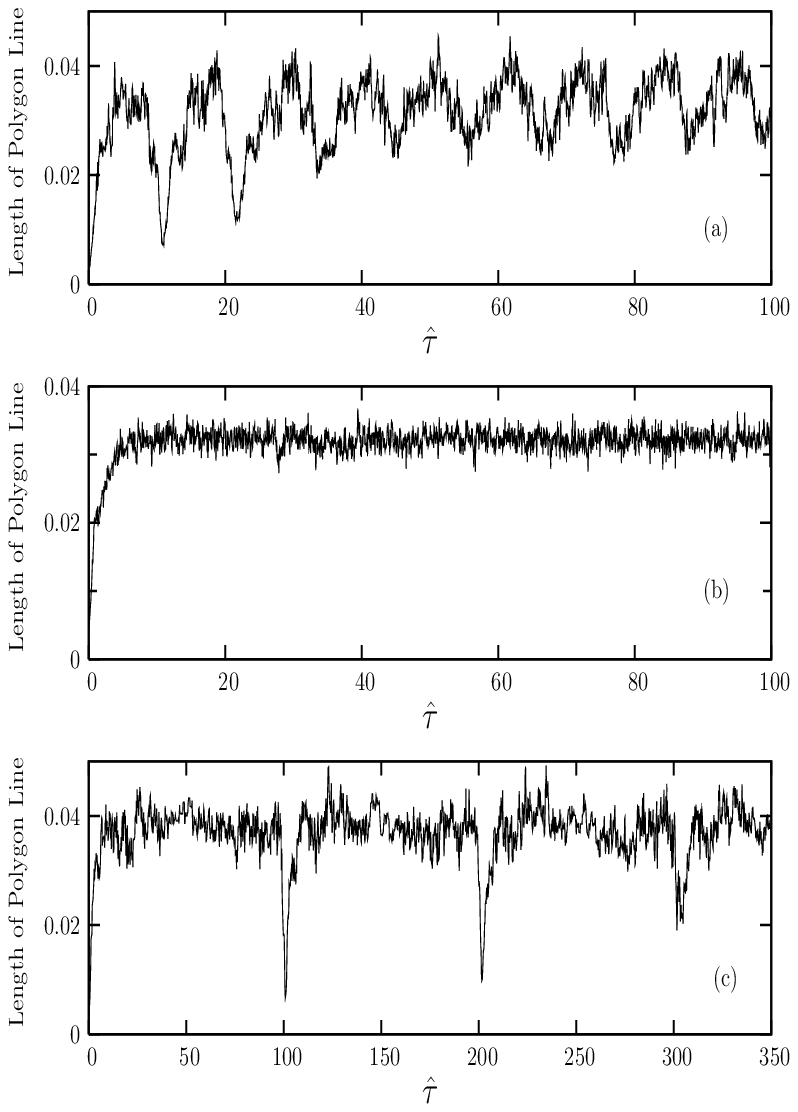}
\caption{\label{lpl}}
\end{figure}

Next, to calculate the length of polygon line \cite{mjbmp1996}, the trajectory 
in $(x,x_{\hat{\tau}},\dot{x})$ space is restricted to a two dimensional
surface. The restriction in dimension is effected by intersecting the projected
trajectory with a surface $k(x,x_{\hat{\tau}},\dot{x})=0$. Consequently the
number of times the trajectory traverses the surface and the corresponding
intersection points can be calculated.  One then orders the points with respect
to the values of $x_{\hat{\tau}}$, and  a simple measure for the alignment of
the points is the length $L$ of polygon line connecting all the ordered
points.  Figure~\ref{lpl}a shows length of polygon line $L$ with constant delay
$\tau_0=10$, where the local minima correspond to the delay time  of the system
we have considered. Figure~\ref{lpl}b shows length of polygon line $L$ with
delay time modulation where there is no remnance of information about delay
time from the trajectory, whereas Fig.~\ref{lpl}c is plotted for $\tau_0=100,
\tau_a=0$, to show that the imprints of delay time carved out in the trajectory
vanishes in Fig.~\ref{lpl}b only due to the delay time modulation and not
because of large delay.

\subsubsection{Average mutual information}
\begin{figure}
\centering
\includegraphics[width=0.6\columnwidth]{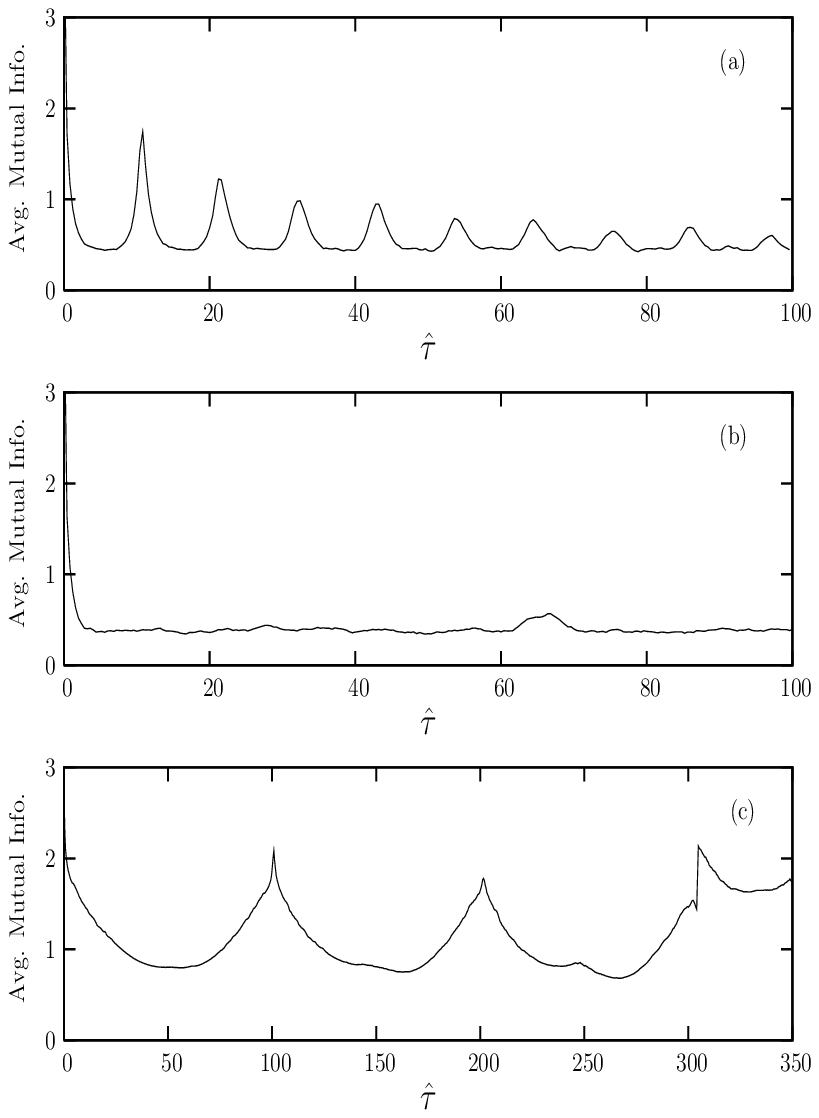}
\caption{\label{ami}}
\end{figure}

As a final example, we will calculate average mutual information defined by
(see for example, \cite{whkmc2004pre,amfhls1986,hdia1996} and references 
therein)
\begin{eqnarray}
I(\hat{\tau})=\sum_{x(n),x(n+\hat{\tau})}P(x(n),x(n+\hat{\tau})) \times
\log_2\left[\frac{P(x(n),x(n+\hat{\tau}))} {P(x(n))P(x(n+\hat{\tau}))}\right],
\label{amieq}
\end{eqnarray}
where $P(x(n),x(n+\hat{\tau}))$ is the joint probability density for
measurements in the chaotic time series $X=(x(1),x(2),...,x(m))$ and in the
constructed delay time series
$X_{\hat{\tau}}=(x(1+\hat{\tau}),x(2+\hat{\tau}),...,x(m+\hat{\tau}))$ by
varying $\hat{\tau}$, resulting in values $x(n)$ and $x(n+\hat{\tau})$.
$P(x(n))$ and $P(x(n+\hat{\tau}))$ are the individual probability densities for
the measurements of $X$ and $X_{\hat{\tau}}$. Figure~\ref{ami} shows the
average mutual information for the cases of constant delay time with
$\tau_0=10$ (Fig.~\ref{ami}a) and with delay time  modulation
(Fig.~\ref{ami}b).  Figure~\ref{ami}c is plotted for $\tau_0=100$  to show that
the absence of local peaks in Fig.~\ref{ami}b is due to delay time modulation
and not because of large delay. For fixed delay time the average mutual
information shows local peaks at the time-delay  $\hat{\tau}=\tau_0$ (or
multiples of it $\hat{\tau}=n\tau_0$) of the system, whereas for the case of
delay time modulation the average mutual information has no such peaks to
identify the delay time of the delayed system.

One can also obtain similar results with other measures such as autocorrelation
function, onestep prediction error and average fitting error
\cite{mjbthm1997,mjbmp1996,czchl1999,vipmdp2002}.  However, we are not
presenting these results here for convenience. In order to perform the phase
space reconstruction, the first step is to find out the delay time for the
projected trajectories.  By introducing the delay time modulation the imprints
of delay time in the projected trajectory is completely removed as seen above
for the present system, inhibiting any possibility of phase space
reconstruction. This is essentially consequent of the fact that when the delay
time is modulated by the fully rectified sine term, the delay time effectively
gets increased in which case the number of positive Lyapunov exponent also
increases (as noted in Fig. 2 in Ref.~\cite{dvskml2005ijbc}. Consequently study
of chaos synchronization in a system of such coupled delay time modulated
oscillators will be of considerable interest.

\section{Coupled system and the stability condition in the presence of delay
time modulation}
Now let us consider the following unidirectionally coupled drive $x_1(t)$ and
response $x_2(t)$ systems with two different modulated time-delays $\tau_1(t)$ and
$\tau_2(t)$ as feedback and coupling time-delays, respectively (hereafter we
write  $\tau_1(t)$  and $\tau_2(t)$ simply as  $\tau_1$ and $\tau_2$
respectively),
\begin{subequations}
\begin{eqnarray}
\dot{x_1}(t)&=&-ax_1(t)+b_{1}f(x_1(t-\tau_{1})),  \\
\dot{x_2}(t)&=&-ax_2(t)+b_{2}f(x_2(t-\tau_{1}))+b_{3}f(x_1(t-\tau_{2})),
\end{eqnarray}
\label{eq.one}
\end{subequations}
where $b_1, b_2$ and $b_3$ are constants, $a>0$, and $f(x)$ is of the same form
as in Eq.~(\ref{eqoneb}) with 
\begin{subequations}
\begin{eqnarray}
\tau_1&=&\tau_{10}+\tau_{1a}\left|\sin(\omega_1 t)\right|,  \\
\tau_2&=&\tau_{20}+\tau_{2a}\left|\sin(\omega_2 t)\right|,
\end{eqnarray}
\label{eq.dtm}
\end{subequations}
where $\tau_{10}$ and $\tau_{20}$ are the zero frequency components of feedback
delay and coupling delay, $\tau_{1a}$ and $\tau_{2a}$ are the amplitudes of the
time dependent components of $\tau_1$ and $\tau_2$, respectively, and  
$\omega_1/\pi$ and $\omega_2/\pi$ are the corresponding frequencies of their 
modulations.

Now we can deduce the stability condition for synchronization of the two
time-delay systems, Eqs.~(\ref{eq.one}a) and (\ref{eq.one}b), in the presence
of the delay coupling  $b_{3}f(x_1(t-\tau_{2}))$ with time delay modulation in
both the feedback delay and coupling delay.  
The time evolution of the difference system with the state variable
$\Delta=x_{1\tau_2-\tau_1}-x_2$, where 
$x_{1\tau_2-\tau_1} = x_{1}(t-(\tau_2-\tau_1))$, can be  written for
small values of $\Delta$ by using the evolution Eqs.~(\ref{eq.one}) as
\begin{align}
\dot{\Delta}=-a\Delta+(b_2+b_3-b_1)f(x_1(t-\tau_2))+b_2f^{\prime}
(x_1(t-\tau_2))\Delta_{\tau_1},\;\;\Delta_\tau=\Delta(t-\tau).  
\end{align}
Then $\Delta=0$ corresponds to anticipatory
synchronization when $\tau_2 < \tau_1$, identical or complete synchronization 
for $\tau_2 = \tau_1$ and lag synchronization when $\tau_2 > \tau_1$.
In order to study the stability of the
synchronization manifold as in the case of constant time delay case
~\cite{dvskml2005}, we choose the parametric condition,
\begin{align}
b_1=b_2+b_3,
\label{eq.paracon}  
\end{align}
so that the evolution equation for the difference system $\Delta$ becomes
\begin{align}
\dot{\Delta}=-a\Delta+b_{2}f^\prime(x_1(t-\tau_{2}))\Delta_{\tau_1}.
\label{eq.difsys}
\end{align}
The synchronization manifold is locally attracting if the origin of this
equation is stable.  Following Krasovskii-Lyapunov functional approach 
\cite{nnk1963,kp1998}, we define a positive definite Lyapunov functional of the
form
\begin{align}
V(t)=\frac{1}{2}\Delta^2+\mu \int_{-\tau_1(t)}^0\Delta^2(t+\theta)d\theta,
\end{align}
where $\mu$  is an arbitrary positive parameter, $\mu>0$.  Note that $V(t)$
approaches zero as $\Delta \rightarrow 0$.

To estimate a sufficient condition for the stability of the solution $\Delta=0$,
we require the derivative of the functional $V(t)$ along the trajectory
of Eq.~(\ref{eq.difsys}),
\begin{align}
\frac{dV}{dt}=-a\Delta^2+b_2f^{\prime}(x_1(t-\tau_2))\Delta
\Delta_{\tau_1}+\mu\left[\Delta^2_{\tau_1}\tau_1^{\prime}+
\Delta^2-\Delta_{\tau_1}^2\right],
\end{align}
to be negative.  Note that in the case of constant modulation 
$\tau_1^{\prime}=\frac{d\tau_1}{dt}$ vanishes. The above equation can be 
rewritten as
\begin{align}
\frac{dV}{dt}&=-\mu\Delta^2\Gamma(X,\mu),
\end{align}
where  $\Gamma=\bigr[\bigr((a-\mu)/\mu\bigl)
-\bigr(b_2f^{\prime}(x_1(t-\tau_2))/\mu\bigl)X+X^2/
(1-\tau_1^{\prime})\bigl]$, $X=\Delta_{\tau_1}/\Delta$.
In order to show that $\frac{dV}{dt}<0$ for all $\Delta$ and $\Delta_\tau$
and so for all $X$, it is sufficient to show that $\Gamma_{min}>0$.
One can easily check that the absolute minimum of $\Gamma$ occurs at
$X=b_2f^{\prime}(x_1(t-\tau_2))/2\mu(1-\tau_1^{\prime})$ with 
$\Gamma_{min}=\bigr[4\mu(a-\mu)(1-\tau_1^{\prime})
-b_2^2f^{\prime 2}
(x_1(t-\tau_2))\bigl]/4\mu^2(1-\tau_1^{\prime})$.
Consequently, we have the condition for stability as
\begin{align}
a>\frac{b_2^2f^{\prime 2}(x_1(t-\tau_2))}{4\mu(1-\tau_1^{\prime})}
+\mu = \Phi(\mu).
\label{eq.ineq}
\end{align}
Again $\Phi(\mu)$ as a function of $\mu$ for a given $f^{\prime}(x)$ has an
absolute minimum at $\mu=\left(\left|\frac{b_2f^{\prime}(x_1(t-\tau_2))}
{2\sqrt{(1-\tau_1^{\prime})}}\right|\right)$ with 
$\Phi_{min}=\left|\frac{b_2f^{\prime}(x_1(t-\tau_2))}{\sqrt{(1-\tau_1^{\prime}})}\right|$.  
Since $\Phi\ge\Phi_{min}=
\left|\frac{b_2f^{\prime}(x_1(t-\tau_2))}{\sqrt{(1-\tau_1^{\prime})}}\right|$, from the 
inequality (\ref{eq.ineq}), it turns out that
the sufficient condition for asymptotic stability is
\begin{align}
a>\left|\frac{b_2f^{\prime}(x_1(t-\tau_2))}{\sqrt{(1-\tau_1^{\prime})}}\right|
\label{eq.asystab}  
\end{align}
along with the condition (\ref{eq.paracon}) on the parameters $b_1,b_2$ and $b_3$.

Now from the form of the piecewise linear function $f(x)$ given by Eq.~(2),
we have,
\begin{align}
\left|f^{\prime}(x_1(t-\tau_2))\right|=
\left\{
\begin{array}{cc}
1.5,&  0.8\leq|x_1|\leq\frac{4}{3}\\
1.0,&  |x_1|<0.8 \\
\end{array} \right.
\end{align}
Consequently the stability condition
(\ref{eq.asystab}) becomes $a>1.5\left|\frac{b_2}
{\sqrt{(1-\tau_1^{\prime})}}\right|>\left|\frac{b_2}
{\sqrt{(1-\tau_1^{\prime})}}\right|$ along with the parametric
restriction $b_1=b_2+b_3$.

Thus one can take $a>\left|\frac{b_2}
{\sqrt{(1-\tau_1^{\prime})}}\right|$ as a less stringent condition for 
(\ref{eq.asystab}) to be valid, while
\begin{align}
a>1.5\left|\frac{b_2}{\sqrt{(1-\tau_1^{\prime})}}\right| 
\label{eq.four}
\end{align}
can be considered as the most general condition specified by (\ref{eq.asystab})
for asymptotic stability of the synchronized state $\Delta=0$.  The condition
(\ref{eq.four}) indeed corresponds to the stability condition for exact
anticipatory/lag as well as exact complete synchronizations for a given value
of the coupling delay $\tau_2$ in a global sense. It may be noted that the
stability condition (\ref{eq.four}) is valid irrespective of the nature of the
coupling delay, that is whether it is constant or modulated. However, when the
feed back delay $\tau_1$ is constant the condition (\ref{eq.four}) reduces to
$a>1.5|b_2|$ as discussed in  ref.~\cite{dvskml2005}. In the following, we will
consider both the possibilities of constant ($\tau_2=\tau_{20}$) and
periodically modulated ($\tau_2=\tau_{20}+\tau_{2a}\left|\sin(\omega_2
t)\right|$) coupling delays with a periodically modulated feedback delay
($\tau_1=\tau_{10}+\tau_{1a}\left|\sin(\omega_1 t)\right|$). We demonstrate
through detailed numerical analysis that  there exists oscillating
synchronization that oscillates between anticipatory, complete and lag
synchronizations  for the case of constant coupling delay $\tau_2=\tau_{20}$.
Intermittent anticipatory/lag and complete synchronizations are shown to exist
for the case of coupling delay with delay time modulation
$\tau_2=\tau_{20}+\tau_{2a}\left|\sin(\omega_2 t)\right|$, when 
$\tau_{2a}=\tau_{1a}$ and $\omega_1=\omega_2$. For $\tau_{2a}\ne\tau_{1a}$  and
$\omega_1\ne\omega_2$, more complicated oscillating type synchronizations
occur.

\section{Oscillating synchronization}

At first we consider the constant coupling delay, $\tau_2=\tau_{20}$, and  show
that there exists oscillating synchronization that oscillates between
anticipatory, complete and lag synchronizations as a function of time for
suitable range of parameters.

Now we will choose the delay time modulation in the form (\ref{eq.dtm}a) for
the feedback delay $\tau_1 (=\tau_{10}+\tau_{1a}\left|\sin(\omega_1 t)\right|)$
with  $\tau_{10}=10, \tau_{1a}=90$ and $\omega_1=10^{-4}$.  We have fixed the
value of $\tau_{2a}=0$ in (\ref{eq.dtm}b), so that the coupling delay becomes
constant $\tau_2=\tau_{20}=45$ with the parameters  $a=1,b_1=1.2$ in
Eq.~(\ref{eq.one}) and the values of $b_2$ and $b_3$ are chosen according to
the  parametric restriction (\ref{eq.paracon}) depending upon the stability
condition to be satisfied.   For the chosen values of $\tau_{10}$ and
$\tau_{1a}$, one can find that  $\tau_1$ oscillates between 
($\tau_1(t)=\tau_{10}+\tau_{1a}\left|\sin(\omega_1
t)\right|=10+90\left|\sin(\omega_1 t)\right|$) the values 10 and 100.   With
the chosen value of constant coupling delay $\tau_2=45$ and time dependent 
feedback delay $\tau_1$, as time evolves one  finds that the feedback delay
$\tau_1(t)$ is lesser than the value of constant coupling delay $\tau_2$
initially for some time  (in which case $\tau(t)=\tau_2-\tau_1(t)>0$, so that
there exists lag synchronization $x_1(t-\tau(t))=x_2(t)$ with varying lag time
$\tau(t)=\tau_2-\tau_1(t)$). As time evolves, $\tau_1(t)$ increases eventually
and it approaches $\tau_1=45$  at a certain later time $(T=\pi/\omega_1)$, where
$\tau(t)=\tau_2-\tau_1(t)=0$, so that $x_1(t)=x_2(t)$ and a  complete
synchronization occurs at a specific value of time. As $\tau_1(t)$ increases
further above the value of $\tau_2=45$,  the delay time $\tau(t)$ becomes
negative, $\tau(t)=\tau_2-\tau_1(t)<0$ with $x_1(t-\tau(t))=x_2(t))$ and there
exists anticipatory synchronization with varying anticipating time
$\tau(t)=\tau_2-\tau_1(t)$. This anticipatory synchronization continues till 
the value of time dependent feed back delay $\tau_1(t)$  decreases to appraoch
the value of the constant coupling delay $\tau_2=10$ after reaching its maximum
value of 100. Therefore as time evolves there is oscillation between lag,
complete and anticipatory synchronizations with time dependent anticipating and
lag times.

\begin{figure}
\begin{center}
\includegraphics[width=1.0\columnwidth]{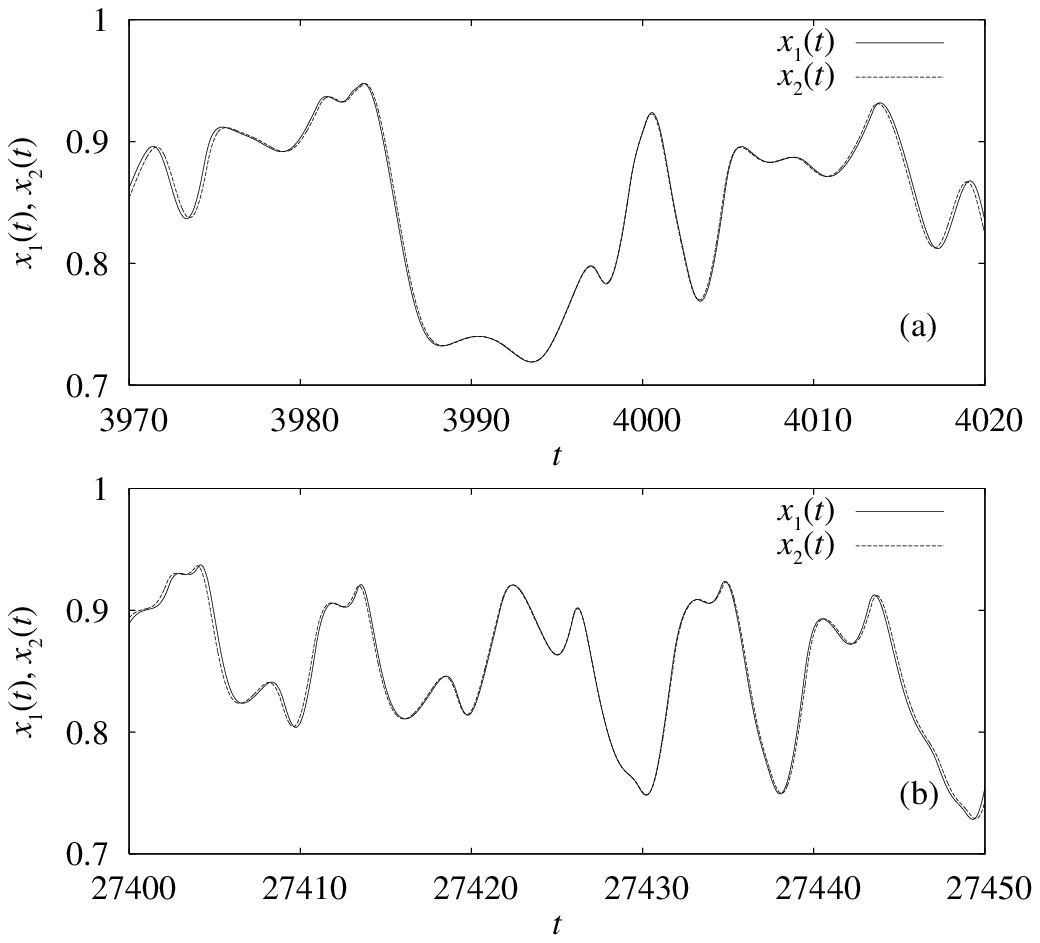}
\end{center}
\caption{\label{osc}} 
\end{figure}

\begin{figure}
\begin{center}
\includegraphics[width=0.8\columnwidth]{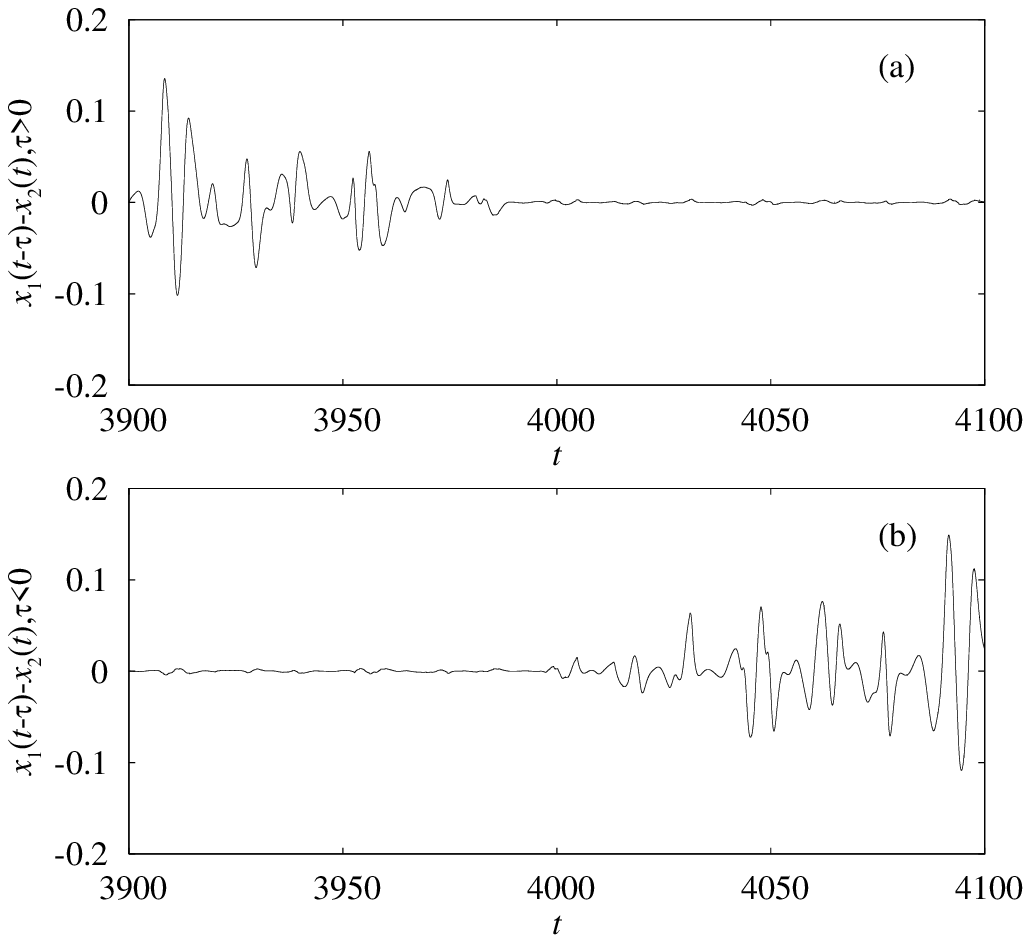}
\end{center}
\caption{\label{dif1}}
\end{figure}

Figure~\ref{osc}a shows the evolution of the drive $x_1(t)$ and the response 
$x_2(t)$ at the transition between lag to anticipatory synchronization via
complete synchronization for the value of  $b_2=0.1$, where the general
stability condition (\ref{eq.four}) is satisfied, whereas  Fig.~\ref{osc}b
shows the evolution of the drive $x_1(t)$ and the response  $x_2(t)$ at the
next transition between anticipatory to lag via complete synchronization. In
Figs.~\ref{dif1}a and~\ref{dif1}b, the difference signals  $x_1(t-\tau)-x_2(t),
\tau>0$ and $x_1(t-\tau)-x_2(t), \tau<0$ are plotted respectively for the value
of parameters satisfying the general stability condition corresponding to the
Fig.~\ref{osc}, confirming the transition between  lag to anticipatory 
synchronization. Thus as a consequence of delay time modulation there exists a
new type of oscillating synchronization that oscillates between anticipatory,
complete and lag synchronizations with varying anticipating and lag times.

\section{Intermittent anticipatory synchronization}
 
Now we consider the coupled time-delay system (\ref{eq.one}) with delay time 
modulation of the  form (\ref{eq.dtm}) in both the feedback and coupling delays
for further studies. We have fixed the values of the parameters as
$a=1,b_1=1.2,\tau_{1a}=\tau_{2a}=90, \omega_1=\omega_2=10^{-5}$ (identical
modulations) and the values  of $b_2$ and $b_3$ are chosen according to the
parametric restriction $b_1=b_2+b_3$ depending upon the stability condition to
be satisfied.  For $\tau_1$, the zero frequency component of amplitude is fixed
as $\tau_{10}=10$ and for $\tau_2$, it is fixed as $\tau_{20}=5$, so that a
constant difference is maintained between  the feedback and the coupling
time delays throughout the time evolution. With the coupling delay $\tau_2
(=5+90\left|\sin(10^{-5} t)\right|)$ being less than the feedback  delay $\tau_1
(=10+90\left|\sin(10^{-5} t)\right|)$, that is $\tau_{2}(t)<\tau_{1}(t)$,  the value of the
anticipating time $\tau=\tau_{2}-\tau_{1}$ turns out to be negative such that
the relation between drive $x_1(t)$ and the response $x_2(t)$ now becomes
$x_1(t-\tau)=x_2(t), \tau<0$, demonstrating  anticipatory synchronization, 
provided the stability condition (\ref{eq.four}) is satisfied with the
parametric restriction specified by Eq.~(\ref{eq.paracon}).

\begin{figure}
\centering
\includegraphics[width=0.7\columnwidth]{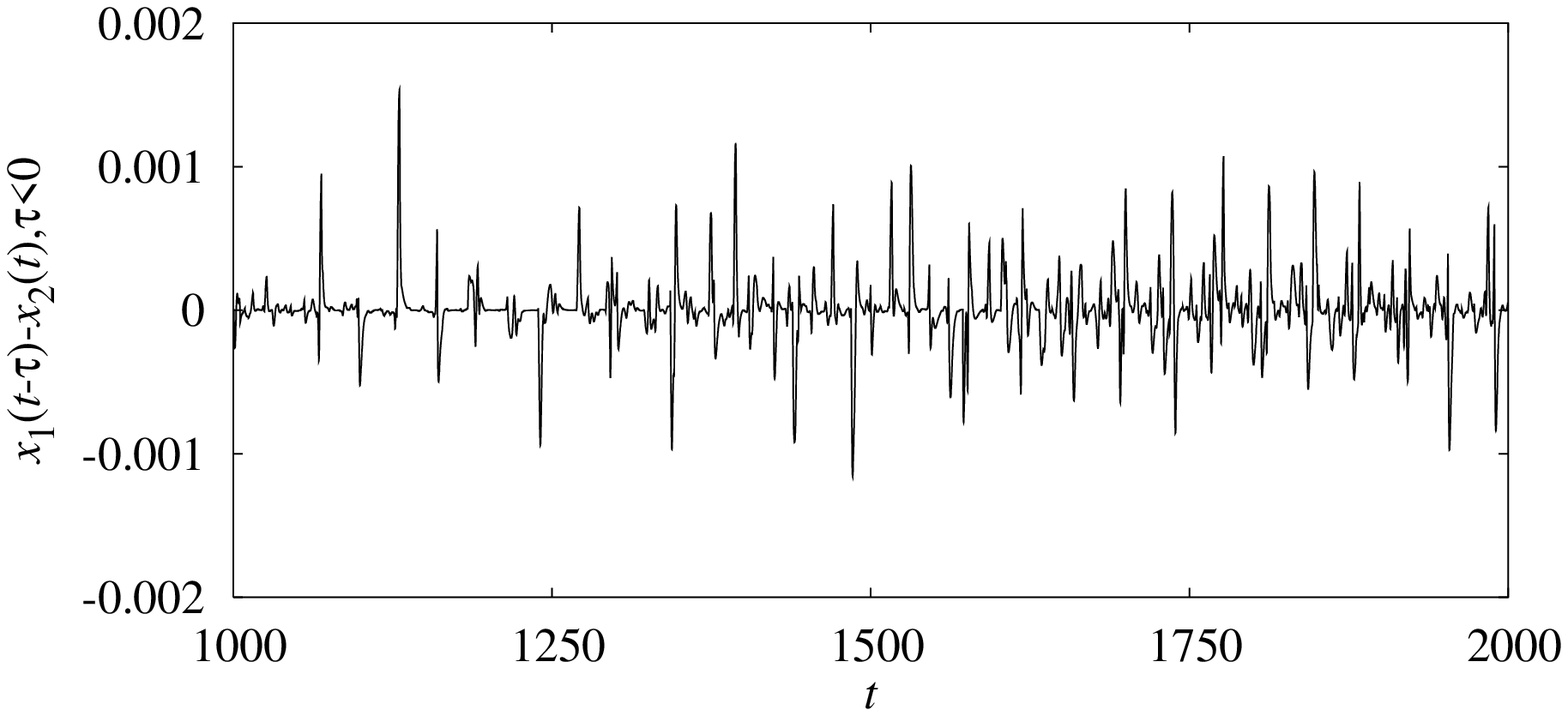}
\caption{\label{apprintts}}
\end{figure}

\begin{figure}
\centering
\includegraphics[width=0.7\columnwidth]{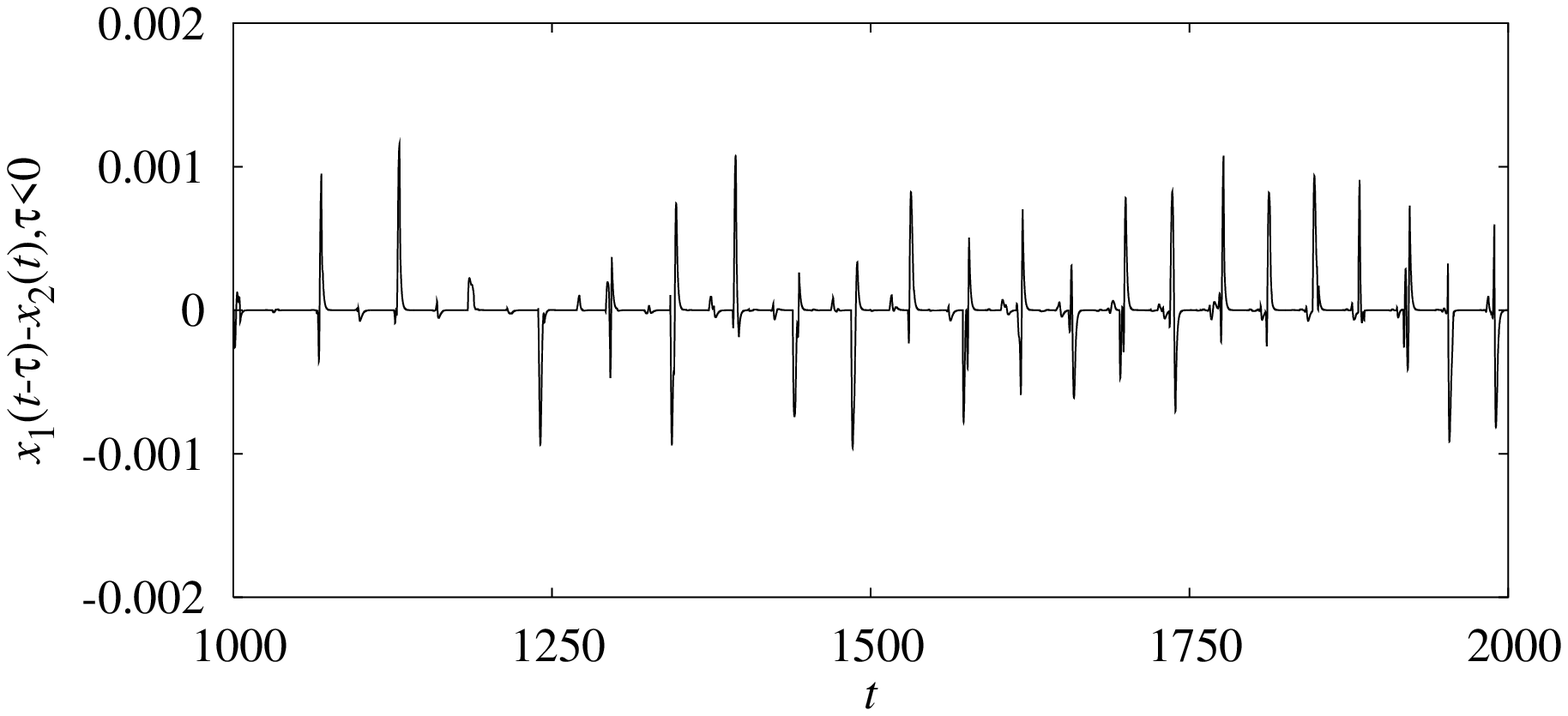}
\caption{\label{intts}}
\end{figure}

Now let us choose the parameter $b_2$ as the control parameter, whose value
determines the stability condition given by Eq.~(\ref{eq.asystab}).
\begin{enumerate}
\item
For $b_2=0 .7$, $1.5\left|\frac{b_2}{\sqrt{1-\tau_1^{\prime}}}\right|>a>
\left|\frac{b_2}{\sqrt{1-\tau_1^{\prime}}}\right|$, the less stringent condition is
satisfied with $\sqrt{1-\tau_1^{\prime}}\approx 1$ for the chosen values of
$\omega$ and $\tau_a$.  One can observe intermittent anticipatory
synchronization as shown in Fig.~\ref{apprintts}, exhibiting typical features
of \emph{on-off} intermittency \cite{npsmh1994,jfhnp1994} with the \emph{off}
state near the laminar phase and the \emph{on} state showing a random burst.
For this value of $b_2$ the amplitude of the laminar phase corresponding to the
synchronized state is approximately zero (of the order $10^{-5}$).
\item
For $b_2=0 .1$, $a>1.5\left|\frac{b_2}{\sqrt{1-\tau_1^{\prime}}}\right|>
\left|\frac{b_2}{\sqrt{1-\tau_1^{\prime}}}\right|$, the general stability condition
(\ref{eq.four}) is satisfied and correspondingly the numerical analysis reveals
that here  the intermittent anticipatory synchronization is such that the
amplitude of the laminar phases corresponding  to the synchronized state is
exactly zero (in the sense that the difference 
$\Delta=x_1(t-\tau)-x_2(t),\tau<0$ is of the order $10^{-16}$ in the laminar
phases) as shown in Fig.~\ref{intts}.
\end{enumerate}

\begin{figure}
\centering
\includegraphics[width=0.5\columnwidth]{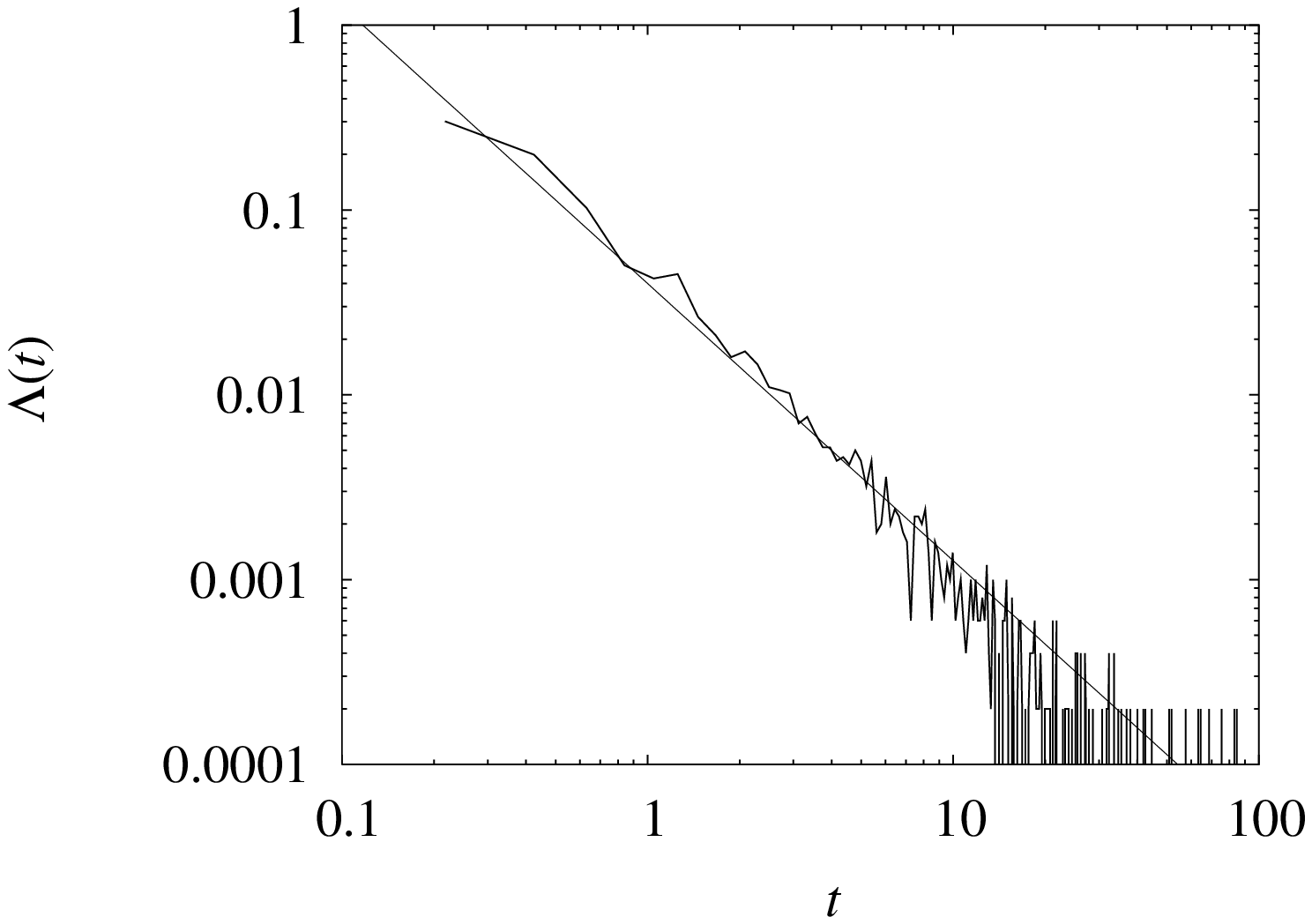}
\caption{\label{apbt}}
\end{figure}

To analyze the statistical features associated with the intermittent nature  
in Fig.~\ref{intts} for the value of $b_2=0.1$,
we have calculated the
distribution of laminar phases $\Lambda(t)$ with the amplitude less than a
threshold value $\Delta <10^{-10}$ and we have observed a universal asymptotic
$-\frac{3}{2}$ power law distribution as shown in Fig.~\ref{apbt}, which is quite
typical for on-off intermittency \cite{npsmh1994,jfhnp1994}.  One can also find
a similar power law distribution for the value of  $b_2=0.7$
discussed above but now with a \emph{larger} threshold 
value ($\Delta <10^{-4}$) of the laminar
region.

\begin{figure}
\centering
\includegraphics[width=0.7\columnwidth]{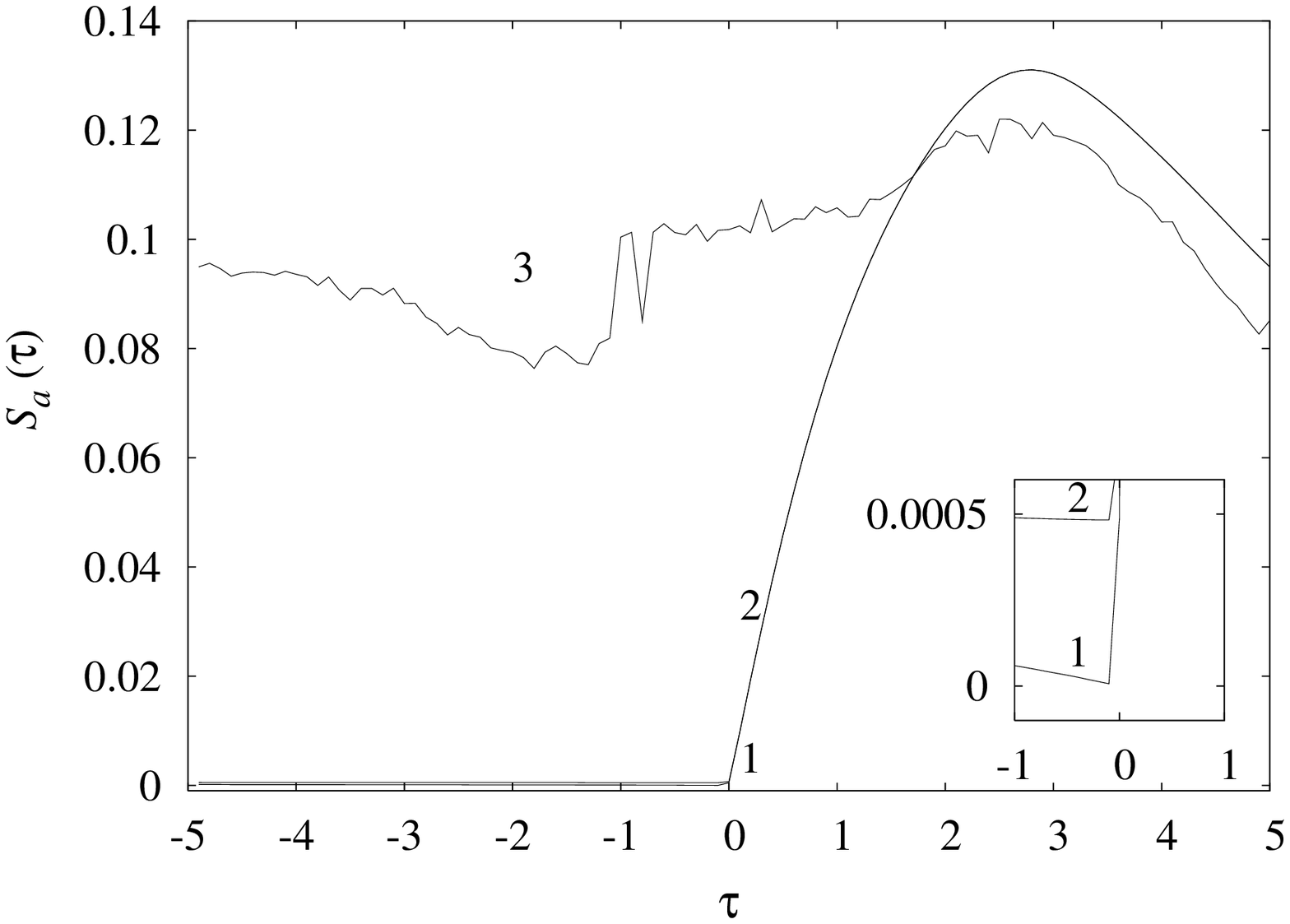}
\caption{\label{aintsim}}
\end{figure}

Now using the notion of similarity function introduced by 
Rosenblum \emph{et al.} \cite{mgrasp1996} to
characterize lag synchronization, one can also characterize anticipatory
synchronization.  Similarity function for anticipatory synchronization is 
defined as the
time-averaged difference between the variables $x_1$ and $x_2$ (with mean
values being subtracted) taken with the  time shift $|\tau|$,
\begin{eqnarray}
S_a^2(\tau)=\frac{\langle[x_1(t+|\tau|)-x_2(t)]^2\rangle}
{[\langle x_{1}^2(t)\rangle\langle x_{2}^2(t)\rangle]^{1/2}},
\label{anti:sim}
\end{eqnarray}
where, $\langle x \rangle$ means time average over the variable $x$. If the
minimum value of $S_a(\tau)$ reaches zero, that is $S_a(\tau)=0$, then there
exists a time shift $|\tau|$ between the two signals $x_1(t)$ and $x_2(t)$ such
that $x_1(t+|\tau|)=x_2(t)$, demonstrating the existence of anticipatory
synchronization between the drive $x_1$ and the response $x_2$ signals.  Figure
~\ref{aintsim} shows the similarity function $S_a(\tau)$ as a function of the
difference between the feedback  and the coupling delays, $\tau=\tau_2-\tau_1$
for three different values of $b_2$, the parameter whose value determines the
stability condition (\ref{eq.asystab}). In Fig.~\ref{aintsim}, the Curve~3 is
plotted for the value of $b_2=1.1$,
$(1.5\left|\frac{b_2}{\sqrt{1-\tau_1^{\prime}}}\right|>
\left|\frac{b_2}{\sqrt{1-\tau_1^{\prime}}}\right|>a)$, where both the less
stringent condition and the most general condition are violated.  From the
curve 3 one can find that the minimum value of $S_a(\tau)$ is greater than zero
for all values of $\tau$, resulting in the lack of exact time shift
(anticipating time) between the drive and the response signals. On the other
hand the curve 2 corresponds to the value of $b_2=0.7$ such that the less
stringent condition is satisfied while the general stability criterion
(\ref{eq.four}) is violated as seen above. Curve 2 shows that the minimum of
similarity function $S_a(\tau)$ is approximately zero  (of the order $10^{-4}$) for
$\tau<0$, as may be seen in the inset of Fig.~\ref{aintsim}, indicating the
existence of  intermittent anticipatory synchronization with the amplitude of
the laminar phases of the difference signal $\Delta=x_1(t-\tau)-x_2(t),\tau<0$,
being approximately zero ($<10^{-5}$). On the other hand, the curve 1 is plotted
for $b_2=0.1$, satisfying the general stability criterion Eq.~(\ref{eq.four}),
which shows that the minimum of  similarity function is much closer to zero (of
order $10^{-8}$), $\tau<0$, indicating that there exists an  intermittent
anticipatory synchronization with the amplitude of the laminar phase of the
difference signal becoming  exactly zero with the anticipating time equal to
the difference between the two time delays $\tau=\tau_2-\tau_1$.

Next, by reducing the value of the modulational frequencies 
$\omega=\omega_1=\omega_2$ further, we find
that the lengths of the laminar phases increase  gradually with a
corresponding  decrease in the number of  turbulent phases. Finally at an
appropriate value of the modulational frequency all the turbulent phases disappear
and there exists only exact anticipatory synchronization without any
intermittent bursts provided the exact stability condition is satisfied.
Correspondingly the similarity function $S_a(\tau)$ becomes zero exactly 
(which is of the order $10^{-16}$) for 
$\tau<0$ in this case, as shown in \cite{dvskml2005}.

\section{Complete synchronization}

\begin{figure}
\centering
\includegraphics[width=0.7\columnwidth]{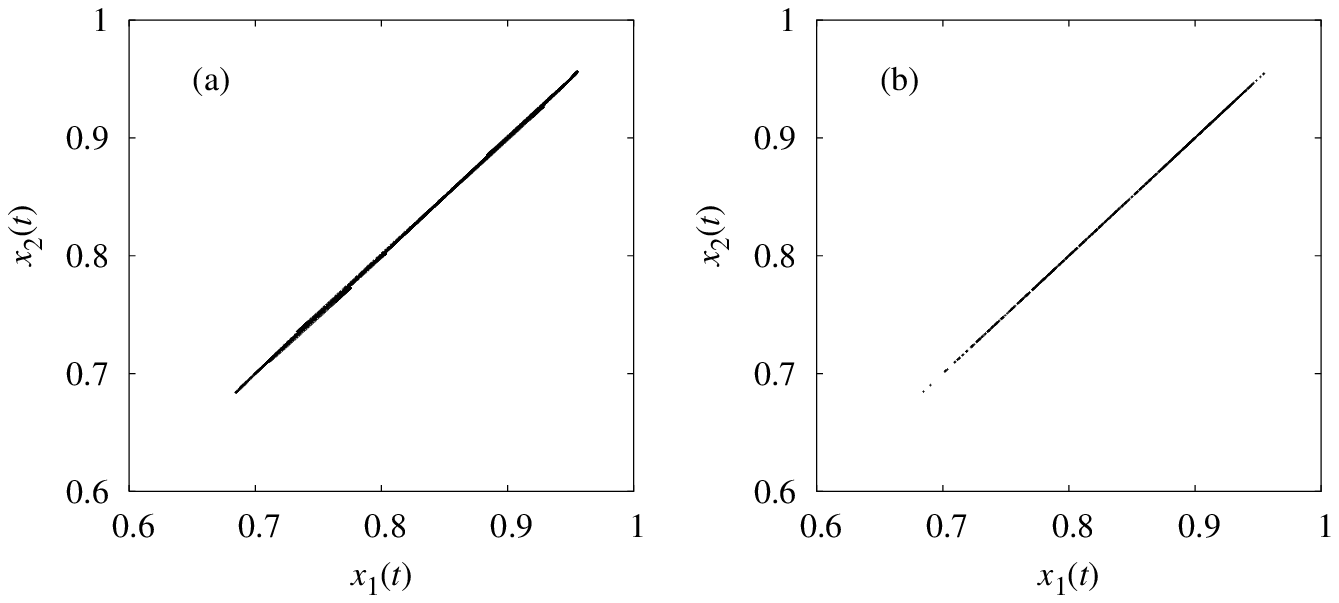}
\caption{\label{cs}}
\end{figure}

Complete synchronization follows the anticipatory synchronization when the
value of the coupling time-delay $\tau_2$ equals the feedback time-delay
$\tau_1$, that is $\tau_{2}=\tau_{1}$, where the anticipating time becomes
$\tau=\tau_{2} - \tau_{1}=0$.   Here also, the same stability criterion
Eq.~(\ref{eq.four}) holds good  with the same parametric restriction specified
by (\ref{eq.paracon}). In this case of complete synchronization
($\tau_{2}=\tau_{1}$), the delay time modulation does not induce any
intermittent nature in the dynamical behavior of the  coupled systems for  any
value of the  modulational frequency $(\omega_1=\omega_2)$ as inferred from
Eq.~(6). Figure~\ref{cs}a shows as  an illustration the plot of $x_1(t)$ vs 
$x_2(t)$ for the values of $b_2=0.7$ and $\omega_1=\omega_2=10^{-5}$, such that
the less stringent condition is satisfied and the general stability criterion
(\ref{eq.four})  is violated. The plot  shows small deviations from the
localized diagonal line implying  an approximate synchronization, whereas
Fig.~\ref{cs}b shows an entirely localized sharp diagonal line for the value of
$b_2=0.1$, where the general stability condition (\ref{eq.four}) is satisfied,
indicating the complete synchronization.

\section{Intermittent lag synchronization}

When the value of the coupling delay $\tau_{2}$ is increased above the value
of the feedback delay $\tau_{1} (\tau_{2}>\tau_{1})$, then the value of the
retarded time $\tau=\tau_{2}-\tau_{1}$ turns out to be positive such that the
relation between the drive $x_1(t)$ and the response $x_2(t)$ now becomes
$x_1(t-\tau)=x_2(t), \tau>0$, depicting the existence of lag synchronization, 
provided
the general stability condition (\ref{eq.four}) is satisfied along with the
parametric condition (\ref{eq.paracon}).

\begin{figure}
\centering
\includegraphics[width=0.7\columnwidth]{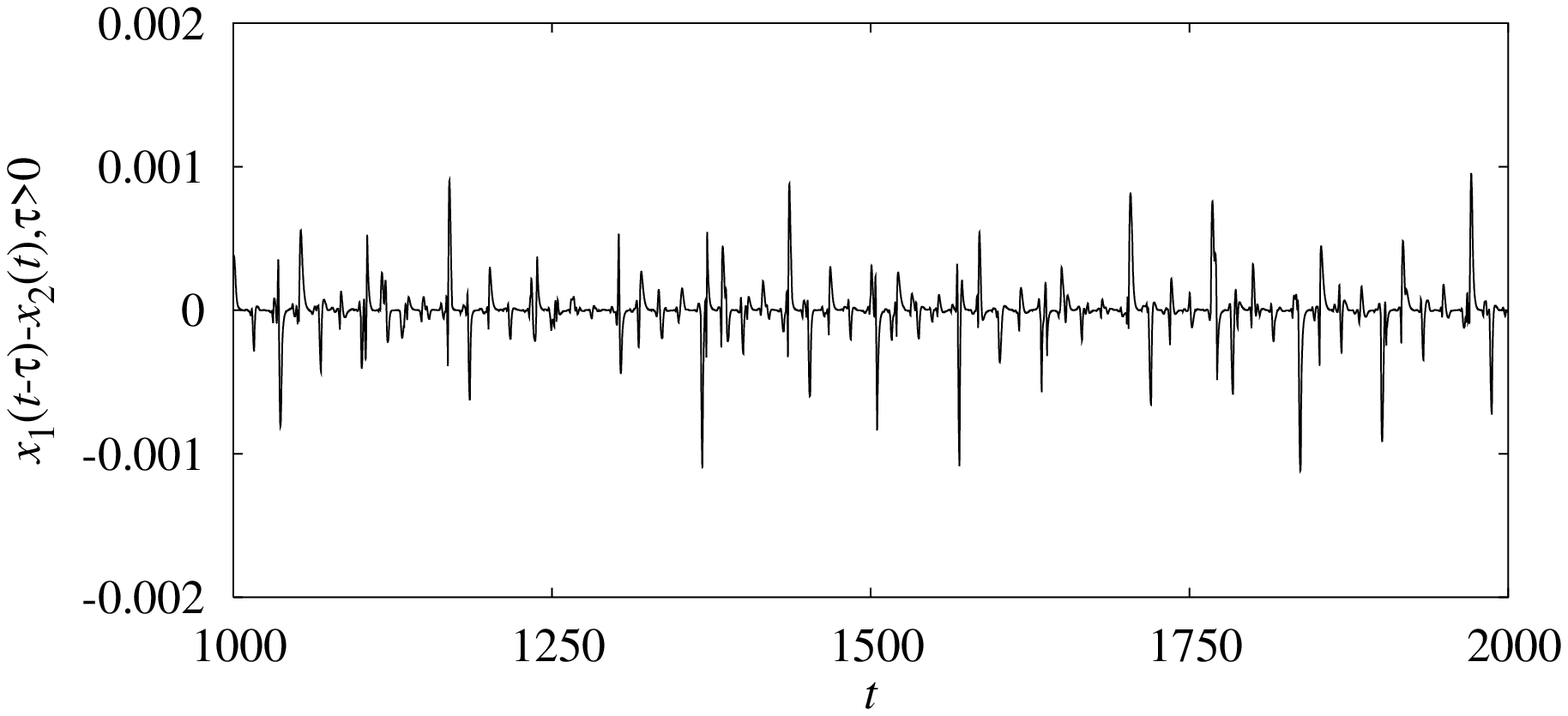}
\caption{\label{apprintlagts}}
\end{figure}

\begin{figure}
\centering
\includegraphics[width=0.7\columnwidth]{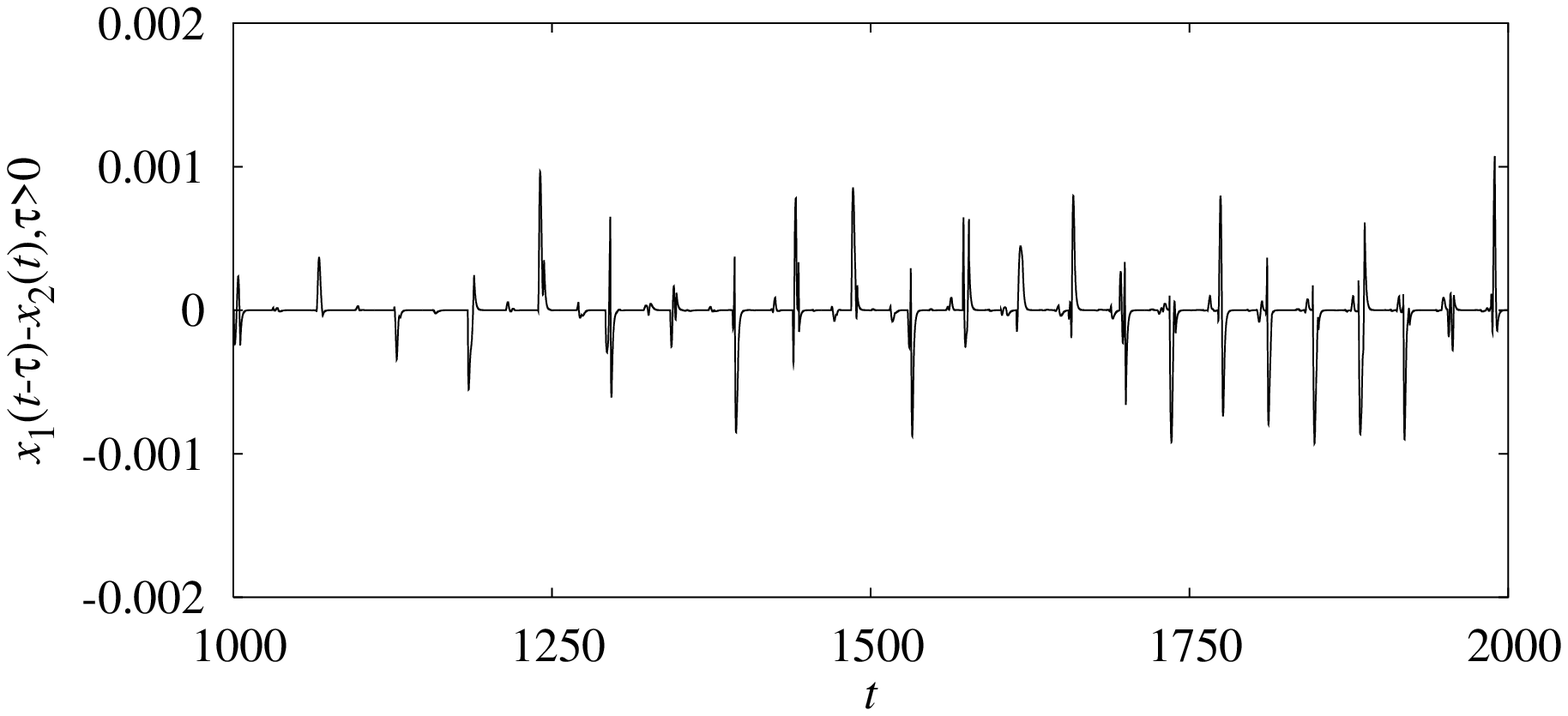}
\caption{\label{lintts}}
\end{figure}

\begin{figure}
\centering
\includegraphics[width=0.5\columnwidth]{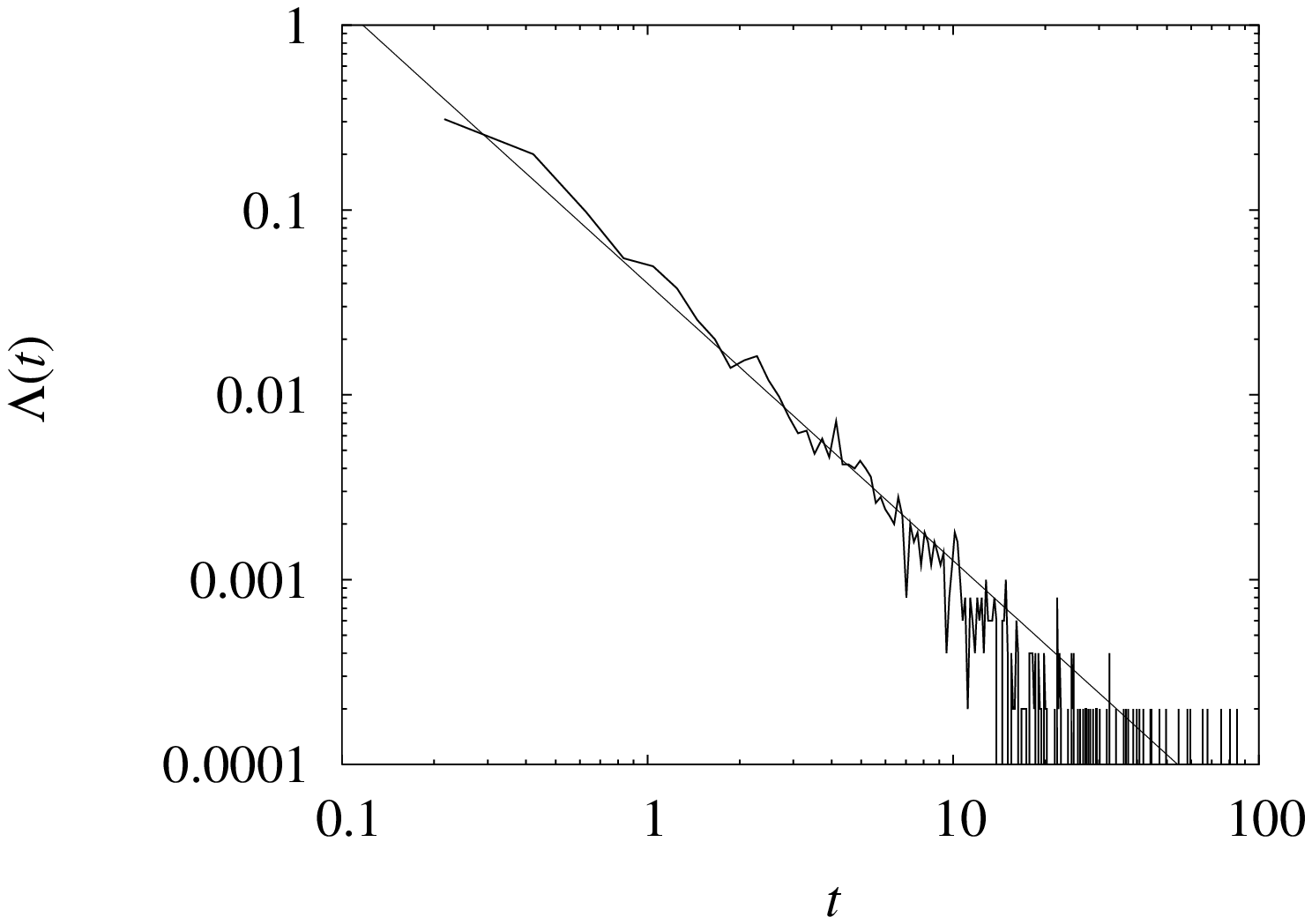}
\caption{\label{lpbt}}
\end{figure}

We have fixed the same values for all the parameters as in the case of
intermittent anticipatory synchronization except for the zero frequency
component $\tau_{20}$ of coupling delay $\tau_2$ which is fixed at 
$\tau_{20}=15$. Figure~\ref{apprintlagts} shows the  intermittent
lag synchronization for the value of $b_2=0.7$, in which case only the less
stringent stability condition is satisfied, where the laminar phase has an
 amplitude which is nearly zero (of the order $10^{-5}$).  Figure~\ref{lintts} shows  intermittent
lag synchronization for the value of $b_2=0.1$, where the amplitude of the
laminar phase vanishes exactly.  In the later case the most
general stability criterion (\ref{eq.four}) is satisfied.  The
statistical behavior associated with the intermittent nature in this case of
intermittent lag synchronization is also characterized by the probability
distribution of laminar phases having amplitude less than a threshold value
$\Delta < 10^{-10}$ corresponding to
a universal asymptotic $-\frac{3}{2}$ power law distribution as shown
in the Fig.~\ref{lpbt}.

The figure shows the probability distribution $\Lambda(t)$ of intermittent lag
synchronization for the value of $b_2=0.1$.  One can also verify that the 
intermittent lag synchronization for the value of $b_2=0.7$ has also similar
power law distribution for \emph{larger} threshold value ($\Delta < 10^{-4}$)of
amplitude of the laminar phases.

\begin{figure}
\centering
\includegraphics[width=0.7\columnwidth]{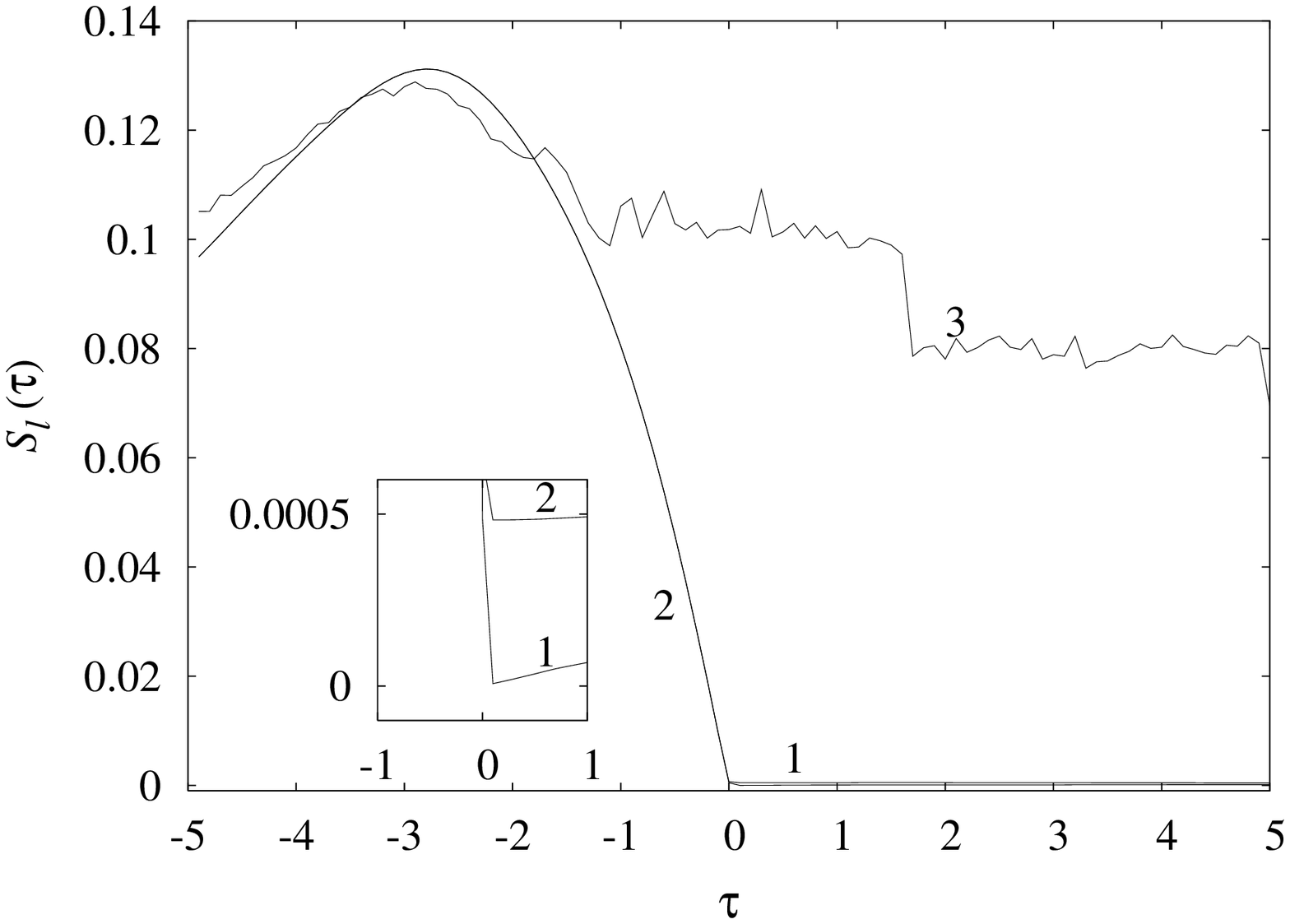}
\caption{\label{lintsim}}
\end{figure}

The existence of intermittent lag synchronization is also characterized by a
similarity function $S_l(\tau)$ defined as
\begin{eqnarray}
S_l^2(\tau)=\frac{\langle[x_1(t-|\tau|)-x_2(t)]^2\rangle}
{[\langle x_{1}^2(t)\rangle\langle x_{2}^2(t)\rangle]^{1/2}}.
\label{lag:sim}
\end{eqnarray}
Figure~\ref{lintsim} shows the similarity function $S_l(\tau)$ for intermittent
lag synchronization as a function of the retarded time $\tau=\tau_2-\tau_1$.
Curve~3 is plotted for the value of $b_2=1.1$ (which is greater than both
$a\sqrt{1-\tau_1^{\prime}}$ and $a\sqrt{1-\tau_1^{\prime}}/1.5$), where the 
minimum of similarity function $S_l(\tau)$ occurs at a finite value of
$S_l(\tau) > 0$ and hence there is a lack of lag synchronization between the 
drive and the response signals indicating asynchronization.  Curve~2
corresponds to the value of $b_2=0.7$, (which is less than
$a\sqrt{1-\tau_1^{\prime}}$ but greater than $a\sqrt{1-\tau_1^{\prime}}/1.5$),
where the minimum of similarity function $S_l(\tau)$ is approximately zero (of
the order of $10^{-4}$, as may be seen in the inset of Fig.~\ref{lintsim})
indicating the existence of intermittent lag synchronization with the amplitude
of the laminar phase being approximately zero. However, for the value of $b_2=0.1$,
for which the general condition (\ref{eq.four}) is obeyed, the minimum of
similarity function for Curve 1 becomes much closer to zero (of the order
$10^{-8}$) which corresponds to intermittent lag synchronization with exact
time shift between the two signals during the laminar phase.

Next, as in the case of intermittent anticipatory synchronization, by reducing
the value of modulational frequency one can find that the lengths of the
laminar phases increase with vanishing turbulent phases and finally at an
appropriate value of the modulational frequency there exists exact lag
synchronization without any intermittent bursts provided the exact stability
condition is satisfied. Correspondingly the similarity function $S_l(\tau)$
becomes zero exactly (which is of the order $10^{-16}$) for  $\tau>0$ in this
case. 

\section{Complex oscillating synchronization}
Finally, when $\tau_{1a}\ne\tau_{2a}$ or/and $\omega_{1}\ne\omega_{2}$ the
frequencies as well as amplitudes of the  modulated feedback delay $\tau_1(t)
(=\tau_{10}+\tau_{1a}\left|\sin(\omega_1 t)\right|)$ and the modulated coupling delay
$\tau_2(t) (=\tau_{20}+\tau_{2a}\left|\sin(\omega_2 t)\right|)$ differ from each other 
resulting in a more complicated variation of the anticipating/lag time
$\tau(t)=\tau_2(t)-\tau_1(t)$. This in turn results in the existence of more
complex oscillating synchronization than the one presented in  Sec.~IV.  It is
clear that one can also introduce  other kinds of modulations instead of
periodic modulation to obtain  varying forms of oscillating synchronizations.

\section{Summary and conclusions}
In this paper, we have shown that there exists a new type of oscillating
synchronization that oscillates between anticipatory, complete and lag
synchronization and vice versa for the case of constant coupling delay with
varying anticipating and lag times.  We have also shown that there exists
regions of intermittent anticipatory synchronization, intermittent lag
synchronization and complete synchronization in the parameter space of $\omega$
and $\tau_2$ with appropriate stability condition for the synchronized state in a
system of two piecewise linear time-delay systems with delay time modulation in
both the feedback and coupling delay.  For a fixed value of $\omega$, we have
shown that there is a transition from intermittent anticipatory to intermittent
lag synchronization through complete  synchronization with the coupling delay
$\tau_2$ as the only control parameter, while all the other parameters are kept
fixed. We have also found that on further reducing the value of $\omega$, one can
observe transition towards exact anticipatory/lag synchronizations from
their intermittent behaviour. The signature of the intermittent behavior in
both the intermittent anticipatory and intermittent lag synchronizations are
characterized by probability distribution of laminar phases satisfying a
universal asymptotic $-\frac{3}{2}$ power law distribution. The existence of
intermittent anticipatory and intermittent lag  synchronizations are
characterized by their corresponding similarity functions. 

Further, we have observed that in the region where the stringent stability
condition (\ref{eq.four}) is satisfied, the minimum of the similarity
functions  $S_a(\tau)$ and $S_l(\tau)$ approaches very closely zero for all
values of $\tau_2 < \tau_1$ and $\tau_2 > \tau_1$, respectively.  The range of
zero values corresponding  to the minimum of similarity functions $S_a(\tau)$
and $S_l(\tau)$ shows the existence of anticipatory and lag synchronizations
for the values of coupling delay $\tau_2$ below and above the feedback delay
$\tau_1$, respectively.   We have also shown that  the estimation of the delay
time carved out of the time series of the delayed system even with delay time 
modulations of fully rectified sinusoidal type is  very difficult by
conventional methods in the present system for suitable choice of the
parameters (in contrast to the chaotic or stochastic delay time modulation as
studied by Kye \emph{et al} \cite{whkmc2004pre} in Mackey-Glass delay system)
and so the messages encoded in such systems can be expected to be less amenable
for extraction by phase space  reconstruction. We have also confirmed
numerically that the phenomena reported in this paper occur in other time-delay
systems such as Mackey-Glass and Ikeda systems also to corroborate the generic
nature of the results.  Also the model system discussed in the present
manuscript is amenable for experimental realization in terms of suitable
nonlinear electronic circuits in view of its piecewise linear nature and we are
pursuing the experimental verification of it.  It is hoped that the study of
such modulated systems will lead to a better understanding of the dynamics of
systems with time dependent topologies such as neural networks, world wide web,
etc.

\section*{Acknowledgments}

This work forms part of a Department of Science and Technology, Government of
India sponsored research project and supported by the Department of Atomic
Energy Commission Raja Ramanna Fellowship to M. L.

\newpage 

\begin{thebibliography}{44}
\expandafter\ifx\csname natexlab\endcsname\relax\def\natexlab#1{#1}\fi
\expandafter\ifx\csname bibnamefont\endcsname\relax
  \def\bibnamefont#1{#1}\fi
\expandafter\ifx\csname bibfnamefont\endcsname\relax
  \def\bibfnamefont#1{#1}\fi
\expandafter\ifx\csname citenamefont\endcsname\relax
  \def\citenamefont#1{#1}\fi
\expandafter\ifx\csname url\endcsname\relax
  \def\url#1{\texttt{#1}}\fi
\expandafter\ifx\csname urlprefix\endcsname\relax\def\urlprefix{URL }\fi
\providecommand{\bibinfo}[2]{#2}
\providecommand{\eprint}[2][]{\url{#2}}

\bibitem[{\citenamefont{Fujisaka and Yamada}(1983{\natexlab{a}})}]{hfty1983}
\bibinfo{author}{\bibfnamefont{H.}~\bibnamefont{Fujisaka}} \bibnamefont{and}
  \bibinfo{author}{\bibfnamefont{T.}~\bibnamefont{Yamada}},
  \bibinfo{journal}{Prog. Theor. Phys.} \textbf{\bibinfo{volume}{69}},
  \bibinfo{pages}{32} (\bibinfo{year}{1983}{\natexlab{a}}).
  
\bibitem[{\citenamefont{Fujisaka and Yamada}(1983{\natexlab{b}})}]{hfty198370}
\bibinfo{author}{\bibfnamefont{H.}~\bibnamefont{Fujisaka}} \bibnamefont{and}
  \bibinfo{author}{\bibfnamefont{T.}~\bibnamefont{Yamada}},
  \bibinfo{journal}{Prog. Theor. Phys.} \textbf{\bibinfo{volume}{70}},
  \bibinfo{pages}{1240} (\bibinfo{year}{1983}{\natexlab{b}}).

\bibitem[{\citenamefont{Pecora and Carroll}(1990)}]{lmptlc1990}
\bibinfo{author}{\bibfnamefont{L.~M.} \bibnamefont{Pecora}} \bibnamefont{and}
  \bibinfo{author}{\bibfnamefont{T.~L.} \bibnamefont{Carroll}},
  \bibinfo{journal}{Phys.\ Rev. Lett.} \textbf{\bibinfo{volume}{64}},
  \bibinfo{pages}{821} (\bibinfo{year}{1990}).

\bibitem[{\citenamefont{Ott et~al.}(1990)\citenamefont{Ott, Gerbogi, and
  Yorke}}]{eocg1990}
\bibinfo{author}{\bibfnamefont{E.}~\bibnamefont{Ott}},
  \bibinfo{author}{\bibfnamefont{C.}~\bibnamefont{Grebogi}}, \bibnamefont{and}
  \bibinfo{author}{\bibfnamefont{J.~A.} \bibnamefont{Yorke}},
  \bibinfo{journal}{Phys. Rev. Lett.} \textbf{\bibinfo{volume}{64}},
  \bibinfo{pages}{1196} (\bibinfo{year}{1990}).

\bibitem[{\citenamefont{Pikovsky et~al.}(2001)\citenamefont{Pikovsky,
  Rosenblum, and Kurths}}]{aspmgr2001}
\bibinfo{author}{\bibfnamefont{A.~S.} \bibnamefont{Pikovsky}},
  \bibinfo{author}{\bibfnamefont{M.~G.} \bibnamefont{Rosenblum}},
  \bibnamefont{and} \bibinfo{author}{\bibfnamefont{J.}~\bibnamefont{Kurths}},
  \emph{\bibinfo{title}{Synchronization - A Unified Approach to Nonlinear
  Science}} (\bibinfo{publisher}{Cambridge University Press},
  \bibinfo{address}{Cambridge}, \bibinfo{year}{2001}).

\bibitem[{\citenamefont{Lakshmanan and Murali}(1996)}]{mlkm1996}
\bibinfo{author}{\bibfnamefont{M.}~\bibnamefont{Lakshmanan}} \bibnamefont{and}
  \bibinfo{author}{\bibfnamefont{K.}~\bibnamefont{Murali}},
  \emph{\bibinfo{title}{Chaos in Nonlinear Oscillators: Controlling and
  Synchronization}} (\bibinfo{publisher}{world Scientific},
  \bibinfo{address}{Singapore}, \bibinfo{year}{1996}).

\bibitem[{\citenamefont{Lakshmanan and Rajasekar}(2003)}]{mlsr2003}
\bibinfo{author}{\bibfnamefont{M.}~\bibnamefont{Lakshmanan}} \bibnamefont{and}
  \bibinfo{author}{\bibfnamefont{S.}~\bibnamefont{Rajasekar}},
  \emph{\bibinfo{title}{Nonlinear Dynamics: Integrability, Chaos and Patterns}}
  (\bibinfo{publisher}{Springer}, \bibinfo{address}{New York},
  \bibinfo{year}{2003}).

\bibitem[{\citenamefont{Rulkov et~al.}(1995)\citenamefont{Rulkov, Sushchik,
  Tsimring, and Abarbanel}}]{nfrmms1995}
\bibinfo{author}{\bibfnamefont{N.~F.} \bibnamefont{Rulkov}},
  \bibinfo{author}{\bibfnamefont{M.~M.} \bibnamefont{Sushchik}},
  \bibinfo{author}{\bibfnamefont{L.~S.} \bibnamefont{Tsimring}},
  \bibnamefont{and} \bibinfo{author}{\bibfnamefont{H.~D.~I.}
  \bibnamefont{Abarbanel}}, \bibinfo{journal}{Phys. Rev. E}
  \textbf{\bibinfo{volume}{51}}, \bibinfo{pages}{980} (\bibinfo{year}{1995}).

\bibitem[{\citenamefont{Brown}(1998)}]{rb1998}
\bibinfo{author}{\bibfnamefont{R.}~\bibnamefont{Brown}},
  \bibinfo{journal}{Phys. Rev. Lett.} \textbf{\bibinfo{volume}{81}},
  \bibinfo{pages}{4835} (\bibinfo{year}{1998}).

\bibitem[{\citenamefont{Kocarev and Parlitz}(1996)}]{lkup1996}
\bibinfo{author}{\bibfnamefont{L.}~\bibnamefont{Kocarev}} \bibnamefont{and}
  \bibinfo{author}{\bibfnamefont{U.}~\bibnamefont{Parlitz}},
  \bibinfo{journal}{Phys. Rev. Lett.} \textbf{\bibinfo{volume}{76}},
  \bibinfo{pages}{1816} (\bibinfo{year}{1996}).

\bibitem[{\citenamefont{Rosenblum et~al.}(1996)\citenamefont{Rosenblum,
  Pikovsky, and Kurths}}]{mgrasp1996}
\bibinfo{author}{\bibfnamefont{M.~G.} \bibnamefont{Rosenblum}},
  \bibinfo{author}{\bibfnamefont{A.~S.} \bibnamefont{Pikovsky}},
  \bibnamefont{and} \bibinfo{author}{\bibfnamefont{J.}~\bibnamefont{Kurths}},
  \bibinfo{journal}{Phys. Rev. Lett.} \textbf{\bibinfo{volume}{76}},
  \bibinfo{pages}{1804} (\bibinfo{year}{1996}).

\bibitem[{\citenamefont{Yalcinkaya and Lai}(1997)}]{tyycl1997}
\bibinfo{author}{\bibfnamefont{T.}~\bibnamefont{Yalcinkaya}} \bibnamefont{and}
  \bibinfo{author}{\bibfnamefont{Y.~C.} \bibnamefont{Lai}},
  \bibinfo{journal}{Phys. Rev. Lett.} \textbf{\bibinfo{volume}{79}},
  \bibinfo{pages}{3885} (\bibinfo{year}{1997}).

\bibitem[{\citenamefont{Rosenblum et~al.}(1997)\citenamefont{Rosenblum,
  Pikovsky, and Kurths}}]{mgrasp1997}
\bibinfo{author}{\bibfnamefont{M.~G.} \bibnamefont{Rosenblum}},
  \bibinfo{author}{\bibfnamefont{A.~S.} \bibnamefont{Pikovsky}},
  \bibnamefont{and} \bibinfo{author}{\bibfnamefont{J.}~\bibnamefont{Kurths}},
  \bibinfo{journal}{Phys. Rev. Lett.} \textbf{\bibinfo{volume}{78}},
  \bibinfo{pages}{4193} (\bibinfo{year}{1997}).

\bibitem[{\citenamefont{Rim et~al.}(2002)\citenamefont{Rim, Kim, Kang, Park,
  and Kim}}]{srik2002}
\bibinfo{author}{\bibfnamefont{S.}~\bibnamefont{Rim}},
  \bibinfo{author}{\bibfnamefont{I.}~\bibnamefont{Kim}},
  \bibinfo{author}{\bibfnamefont{P.}~\bibnamefont{Kang}},
  \bibinfo{author}{\bibfnamefont{Y.~J.} \bibnamefont{Park}}, \bibnamefont{and}
  \bibinfo{author}{\bibfnamefont{C.~M.} \bibnamefont{Kim}},
  \bibinfo{journal}{Phys. Rev. E} \textbf{\bibinfo{volume}{66}},
  \bibinfo{pages}{015205(R)} (\bibinfo{year}{2002}).

\bibitem[{\citenamefont{Zhan et~al.}(2002)\citenamefont{Zhan, Wei, and
  Lai}}]{mzgww2002}
\bibinfo{author}{\bibfnamefont{M.}~\bibnamefont{Zhan}},
  \bibinfo{author}{\bibfnamefont{G.~W.} \bibnamefont{Wei}}, \bibnamefont{and}
  \bibinfo{author}{\bibfnamefont{C.~H.} \bibnamefont{Lai}},
  \bibinfo{journal}{Phys. Rev. E} \textbf{\bibinfo{volume}{65}},
  \bibinfo{pages}{036202} (\bibinfo{year}{2002}).

\bibitem[{\citenamefont{Voss}(2002)}]{huv2002}
\bibinfo{author}{\bibfnamefont{H.~U.} \bibnamefont{Voss}},
  \bibinfo{journal}{Phys. Rev. E} \textbf{\bibinfo{volume}{61}},
  \bibinfo{pages}{5115} (\bibinfo{year}{2002}).

\bibitem[{\citenamefont{Voss}(2001)}]{huv2001}
\bibinfo{author}{\bibfnamefont{H.~U.} \bibnamefont{Voss}},
  \bibinfo{journal}{Phys. Rev. Lett.} \textbf{\bibinfo{volume}{87}},
  \bibinfo{pages}{014102} (\bibinfo{year}{2001}).

\bibitem[{\citenamefont{Masoller}(2001)}]{cm2001}
\bibinfo{author}{\bibfnamefont{C.}~\bibnamefont{Masoller}},
  \bibinfo{journal}{Phys. Rev. Lett.} \textbf{\bibinfo{volume}{86}},
  \bibinfo{pages}{2782} (\bibinfo{year}{2001}).

\bibitem[{\citenamefont{Sivaprakasam et~al.}(2001)\citenamefont{Sivaprakasam,
  Shahverdiev, and Shore}}]{ssems2001}
\bibinfo{author}{\bibfnamefont{S.}~\bibnamefont{Sivaprakasam}},
  \bibinfo{author}{\bibfnamefont{E.~M.} \bibnamefont{Shahverdiev}},
  \bibinfo{author}{\bibfnamefont{P.~S.} \bibnamefont{Spencer}},
  \bibnamefont{and} \bibinfo{author}{\bibfnamefont{K.~A.} \bibnamefont{Shore}},
  \bibinfo{journal}{Phys. Rev. Lett.} \textbf{\bibinfo{volume}{87}},
  \bibinfo{pages}{154101} (\bibinfo{year}{2001}).

\bibitem[{\citenamefont{Shahverdiev et~al.}(2002)\citenamefont{Shahverdiev,
  Sivaprakasam, and Shore}}]{emsss2002}
\bibinfo{author}{\bibfnamefont{E.~M.} \bibnamefont{Shahverdiev}},
  \bibinfo{author}{\bibfnamefont{S.}~\bibnamefont{Sivaprakasam}},
  \bibnamefont{and} \bibinfo{author}{\bibfnamefont{K.~A.} \bibnamefont{Shore}},
  \bibinfo{journal}{Phys. Rev. E} \textbf{\bibinfo{volume}{66}},
  \bibinfo{pages}{017206} (\bibinfo{year}{2002}).


\bibitem[{\citenamefont{Boccaletti and Valladares}(2000)}]{sbdlv2000}
\bibinfo{author}{\bibfnamefont{S.}~\bibnamefont{Boccaletti}} \bibnamefont{and}
  \bibinfo{author}{\bibfnamefont{D.~L.} \bibnamefont{Valladares}},
  \bibinfo{journal}{Phys. Rev. E} \textbf{\bibinfo{volume}{62}},
  \bibinfo{pages}{7497} (\bibinfo{year}{2000}).

\bibitem[{\citenamefont{Valladares and Boccaletti}(2001)}]{dlvsb2001}
\bibinfo{author}{\bibfnamefont{D.~L.} \bibnamefont{Valladares}}
  \bibnamefont{and}
  \bibinfo{author}{\bibfnamefont{S.}~\bibnamefont{Boccaletti}},
  \bibinfo{journal}{Int. J. Bifurcation and Chaos}
  \textbf{\bibinfo{volume}{11}}, \bibinfo{pages}{2699} (\bibinfo{year}{2001}).

\bibitem[{\citenamefont{Taherion and Lai}(1999)}]{stycl1999}
\bibinfo{author}{\bibfnamefont{S.}~\bibnamefont{Taherion}} \bibnamefont{and}
  \bibinfo{author}{\bibfnamefont{Y.~C.} \bibnamefont{Lai}},
  \bibinfo{journal}{Phys. Rev. E} \textbf{\bibinfo{volume}{59}},
  \bibinfo{pages}{R6247} (\bibinfo{year}{1999}).

\bibitem[{\citenamefont{Fischer et~al.}(2006)\citenamefont{Fischer,
  Vicete, Buldu, Peil, Mirasso, Torrent, and Ojalvo}}]{ifrv2006}
\bibinfo{author}{\bibfnamefont{Ingo}~\bibnamefont{Fischer}},
  \bibinfo{author}{\bibfnamefont{Raul} \bibnamefont{Vicente}},
  \bibinfo{author}{\bibfnamefont{Javier M.} \bibnamefont{Buldu}},
  \bibinfo{author}{\bibfnamefont{Michael} \bibnamefont{Peil}},
  \bibinfo{author}{\bibfnamefont{Claudio R.} \bibnamefont{Mirasso}},
  \bibinfo{author}{\bibfnamefont{M.~C.~} \bibnamefont{Torrent}},
  \bibnamefont{and} \bibinfo{author}{\bibfnamefont{Jordi}
  \bibnamefont{Garcia-Ojalvo}},
  \bibinfo{journal}{Phys. Rev. Lett.} \textbf{\bibinfo{volume}{97}},
  \bibinfo{pages}{123902} (\bibinfo{year}{2006}).

\bibitem[{\citenamefont{Kye et~al.}(2004{\natexlab{b}})\citenamefont{Kye, Choi,
  Rim, Kurdoglyan, Kim, and Park}}]{whkmc2004pre}
\bibinfo{author}{\bibfnamefont{W.~H.} \bibnamefont{Kye}},
  \bibinfo{author}{\bibfnamefont{M.}~\bibnamefont{Choi}},
  \bibinfo{author}{\bibfnamefont{S.}~\bibnamefont{Rim}},
  \bibinfo{author}{\bibfnamefont{M.~S.} \bibnamefont{Kurdoglyan}},
  \bibinfo{author}{\bibfnamefont{C.~M.} \bibnamefont{Kim}}, \bibnamefont{and}
  \bibinfo{author}{\bibfnamefont{Y.~J.} \bibnamefont{Park}},
  \bibinfo{journal}{Phys. Rev. E} \textbf{\bibinfo{volume}{69}},
  \bibinfo{pages}{055202(R)} (\bibinfo{year}{2004}{\natexlab{b}}).

\bibitem[{\citenamefont{Atay}(2003)}]{fma2003}
\bibinfo{author}{\bibfnamefont{F.~M.}~\bibnamefont{Atay}},
  \bibinfo{journal}{Phys. Rev. Lett} \textbf{\bibinfo{volume}{91}},
  \bibinfo{pages}{094101} (\bibinfo{year}{2003}).
 

\bibitem[{\citenamefont{Kye et~al.}(2004{\natexlab{a}})\citenamefont{Kye, Choi,
  kim, Lee, Rim, Kim, and Park}}]{whkmc2004}
\bibinfo{author}{\bibfnamefont{W.~H.} \bibnamefont{Kye}},
  \bibinfo{author}{\bibfnamefont{M.}~\bibnamefont{Choi}},
  \bibinfo{author}{\bibfnamefont{M.~W.} \bibnamefont{kim}},
  \bibinfo{author}{\bibfnamefont{S.~Y.} \bibnamefont{Lee}},
  \bibinfo{author}{\bibfnamefont{S.}~\bibnamefont{Rim}},
  \bibinfo{author}{\bibfnamefont{C.~M.} \bibnamefont{Kim}}, \bibnamefont{and}
  \bibinfo{author}{\bibfnamefont{Y.~J.} \bibnamefont{Park}},
  \bibinfo{journal}{Phys. Lett. A} \textbf{\bibinfo{volume}{322}},
  \bibinfo{pages}{338} (\bibinfo{year}{2004}{\natexlab{a}}).
  
\bibitem[{\citenamefont{Kye et~al.}(2004{\natexlab{b}})\citenamefont{Kye, Choi,
  and Kim}}]{whkmc2005pre}
\bibinfo{author}{\bibfnamefont{W.~H.} \bibnamefont{Kye}},
  \bibinfo{author}{\bibfnamefont{M.}~\bibnamefont{Choi}}, 
  \bibinfo{author}{\bibfnamefont{C.~M.} \bibnamefont{Kim}}, \bibnamefont{and}
  \bibinfo{author}{\bibfnamefont{Y.~J.} \bibnamefont{Park}}, 
  \bibinfo{journal}{Phys. Rev. E} \textbf{\bibinfo{volume}{71}},
  \bibinfo{pages}{045202(R)} (\bibinfo{year}{2005}{\natexlab{b}}).
  
\bibitem[{\citenamefont{Shahverdiev and Shore}(2005)}]{emskas2005}
\bibinfo{author}{\bibfnamefont{E.~M.} \bibnamefont{Shahverdiev}}
  \bibnamefont{and} \bibinfo{author}{\bibfnamefont{K.~A.} \bibnamefont{Shore}},
  \bibinfo{journal}{Phys. Rev. E} \textbf{\bibinfo{volume}{71}},
  \bibinfo{pages}{016201} (\bibinfo{year}{2005}).

\bibitem[{\citenamefont{Shahverdiev et~al.}(2004)\citenamefont{Shahverdiev,
  Nuriev, Hashimov, and Shore}}]{nlinemskas2005}
\bibinfo{author}{\bibfnamefont{E.~M.} \bibnamefont{Shahverdiev}},
  \bibinfo{author}{\bibfnamefont{R.~A.} \bibnamefont{Nuriev}},
  \bibinfo{author}{\bibfnamefont{R.~H.} \bibnamefont{Hashimov}},
  \bibnamefont{and} \bibinfo{author}{\bibfnamefont{K.~A.} \bibnamefont{Shore}},
  \bibinfo{journal}{nlin.CD/0404050}  (\bibinfo{year}{2004}).

\bibitem[{\citenamefont{Senthilkumar and
  Lakshmanan}(2005{\natexlab{a}})}]{dvskml2005}
\bibinfo{author}{\bibfnamefont{D.~V.} \bibnamefont{Senthilkumar}}
  \bibnamefont{and}
  \bibinfo{author}{\bibfnamefont{M.}~\bibnamefont{Lakshmanan}},
  \bibinfo{journal}{Phys. Rev. E} \textbf{\bibinfo{volume}{71}},
  \bibinfo{pages}{016211} (\bibinfo{year}{2005}{\natexlab{a}}).

\bibitem[{\citenamefont{Platt et~al.}(1994)\citenamefont{Platt, Hammel, and
  Heagy}}]{npsmh1994}
\bibinfo{author}{\bibfnamefont{N.}~\bibnamefont{Platt}},
  \bibinfo{author}{\bibfnamefont{S.~M.} \bibnamefont{Hammel}},
  \bibnamefont{and} \bibinfo{author}{\bibfnamefont{J.~F.} \bibnamefont{Heagy}},
  \bibinfo{journal}{Phys. Rev. Lett} \textbf{\bibinfo{volume}{72}},
  \bibinfo{pages}{3498} (\bibinfo{year}{1994}).

\bibitem[{\citenamefont{Heagy et~al.}(1994)\citenamefont{Heagy, Platt, and
  Hammel}}]{jfhnp1994}
\bibinfo{author}{\bibfnamefont{J.~F.} \bibnamefont{Heagy}},
  \bibinfo{author}{\bibfnamefont{N.}~\bibnamefont{Platt}}, \bibnamefont{and}
  \bibinfo{author}{\bibfnamefont{S.~M.} \bibnamefont{Hammel}},
  \bibinfo{journal}{Phys. Rev. E} \textbf{\bibinfo{volume}{49}},
  \bibinfo{pages}{1140} (\bibinfo{year}{1994}).

\bibitem[{\citenamefont{Thangavel et~al.}(1998)\citenamefont{Thangavel, Murali,
  and Lakshmanan}}]{ptkm1998}
\bibinfo{author}{\bibfnamefont{P.}~\bibnamefont{Thangavel}},
  \bibinfo{author}{\bibfnamefont{K.}~\bibnamefont{Murali}}, \bibnamefont{and}
  \bibinfo{author}{\bibfnamefont{M.}~\bibnamefont{Lakshmanan}},
  \bibinfo{journal}{Int. J. Bifurcation and Chaos}
  \textbf{\bibinfo{volume}{8}}, \bibinfo{pages}{2481} (\bibinfo{year}{1998}).

\bibitem[{\citenamefont{Senthilkumar and
  Lakshmanan}(2005{\natexlab{b}})}]{dvskml2005ijbc}
\bibinfo{author}{\bibfnamefont{D.~V.} \bibnamefont{Senthilkumar}}
  \bibnamefont{and}
  \bibinfo{author}{\bibfnamefont{M.}~\bibnamefont{Lakshmanan}},
  \bibinfo{journal}{Int. J. Bifurcation and Chaos}
  \textbf{\bibinfo{volume}{15}}, \bibinfo{pages}{2895} 
  (\bibinfo{year}{2005}{\natexlab{b}}).
  
\bibitem[{\citenamefont{Lu and He}(1996)}]{hlzh1996}
\bibinfo{author}{\bibfnamefont{H.}~\bibnamefont{Lu}} \bibnamefont{and}
  \bibinfo{author}{\bibfnamefont{Z.}~\bibnamefont{He}}, \bibinfo{journal}{IEEE
  Trans. Circuits Syst. I} \textbf{\bibinfo{volume}{43}}, \bibinfo{pages}{700}
  (\bibinfo{year}{1996}).

\bibitem[{\citenamefont{Bunner et~al.}(1997)\citenamefont{Bunner, Meyer,
  Kittel, and Parisi}}]{mjbthm1997}
\bibinfo{author}{\bibfnamefont{M.~J.} \bibnamefont{Bunner}},
  \bibinfo{author}{\bibfnamefont{T.}~\bibnamefont{Meyer}},
  \bibinfo{author}{\bibfnamefont{A.}~\bibnamefont{Kittel}}, \bibnamefont{and}
  \bibinfo{author}{\bibfnamefont{J.}~\bibnamefont{Parisi}},
  \bibinfo{journal}{Phys. Rev. E} \textbf{\bibinfo{volume}{56}},
  \bibinfo{pages}{5083} (\bibinfo{year}{1997}).

\bibitem[{\citenamefont{Bunner et~al.}(1996)\citenamefont{Bunner, Popp, Meyer,
  Kittel, and Parisi}}]{mjbmp1996}
\bibinfo{author}{\bibfnamefont{M.~J.} \bibnamefont{Bunner}},
  \bibinfo{author}{\bibfnamefont{M.}~\bibnamefont{Popp}},
  \bibinfo{author}{\bibfnamefont{T.}~\bibnamefont{Meyer}},
  \bibinfo{author}{\bibfnamefont{A.}~\bibnamefont{Kittel}}, \bibnamefont{and}
  \bibinfo{author}{\bibfnamefont{J.}~\bibnamefont{Parisi}},
  \bibinfo{journal}{Phys. Rev. E} \textbf{\bibinfo{volume}{54}},
  \bibinfo{pages}{R3082} (\bibinfo{year}{1996}).
  

\bibitem[{\citenamefont{Fraser and Swinney}(1986)}]{amfhls1986}
\bibinfo{author}{\bibfnamefont{A.~M.} \bibnamefont{Fraser}} \bibnamefont{and}
  \bibinfo{author}{\bibfnamefont{H.~L.} \bibnamefont{Swinney}},
  \bibinfo{journal}{Phys. Rev. A} \textbf{\bibinfo{volume}{33}},
  \bibinfo{pages}{1134} (\bibinfo{year}{1986}).

\bibitem[{\citenamefont{Abarbanel}(1996)}]{hdia1996}
\bibinfo{author}{\bibfnamefont{H.~D.~I.} \bibnamefont{Abarbanel}},
  \emph{\bibinfo{title}{Analysis of Observed Chaotic Data}}
  (\bibinfo{publisher}{Springer}, \bibinfo{address}{Newyork},
  \bibinfo{year}{1996}).
  
\bibitem[{\citenamefont{Zhou and Lai}(1999)}]{czchl1999}
\bibinfo{author}{\bibfnamefont{C.}~\bibnamefont{Zhou}} \bibnamefont{and}
  \bibinfo{author}{\bibfnamefont{C.~H.} \bibnamefont{Lai}},
  \bibinfo{journal}{Phys. Rev. E} \textbf{\bibinfo{volume}{60}},
  \bibinfo{pages}{R320} (\bibinfo{year}{1999}).

\bibitem[{\citenamefont{Ponomarenko and Prokhorov}(2002)}]{vipmdp2002}
\bibinfo{author}{\bibfnamefont{V.~I.} \bibnamefont{Ponomarenko}}
  \bibnamefont{and} \bibinfo{author}{\bibfnamefont{M.~D.}
  \bibnamefont{Prokhorov}}, \bibinfo{journal}{Phys. Rev. E}
  \textbf{\bibinfo{volume}{66}}, \bibinfo{pages}{026215}
  (\bibinfo{year}{2002}).


\bibitem[{\citenamefont{Krasovskii}(1963)}]{nnk1963}
\bibinfo{author}{\bibfnamefont{N.~N.} \bibnamefont{Krasovskii}},
  \emph{\bibinfo{title}{Stability of Motion}} (\bibinfo{publisher}{Stanford
  University Press}, \bibinfo{address}{Stanford}, \bibinfo{year}{1963}).

\bibitem[{\citenamefont{Pyragas}(1998)}]{kp1998}
\bibinfo{author}{\bibfnamefont{K.}~\bibnamefont{Pyragas}},
  \bibinfo{journal}{Phys. Rev. E} \textbf{\bibinfo{volume}{58}},
  \bibinfo{pages}{3067} (\bibinfo{year}{1998}).

\end{thebibliography}

\newpage
\section*{Figure captions}
\begin{enumerate}
\item Filling factor as a function of delay time $\hat{\tau}$ (of
delayed time series $x_{\hat{\tau}}$). (a) with constant delay $\tau_0=10$ when 
$\tau_a=0$, (b) with delay time modulation of the form (\ref{eqonec}) with
$\tau_0=10, \tau_a=90$ and $\omega=10^{-4}$ and (c) with large constant delay
$\tau_0=100 (\tau_a=0)$.

\item Length of polygon line as a function of delay time
$\hat{\tau}$ (of delayed time series $x_{\hat{\tau}}$). (a) with constant delay
$\tau_0=10 (\tau_a=0)$, (b) with delay time modulation of the form
(\ref{eqonec}) with the same parameters as in  Fig.~\ref{ff} and (c) with large
constant delay $\tau_0=100 (\tau_a=0)$.

\item  Average mutual information as a function of delay time
$\hat{\tau}$ (of delayed time series $x_{\hat{\tau}}$). (a) with constant delay
$\tau_0=10, \tau_a=0$, (b) with delay time modulation of the form
(\ref{eqonec}) with the same parameters as in  Fig.~\ref{ff} and (c) with large
constant delay $\tau_0=100, \tau_a=0$.

\item Oscillating synchronization  for the constant coupling delay
$\tau_2=45$ with time dependent feedback delay of the form (\ref{eq.dtm}a) with
$\tau_{10}=10, \tau_{1a}=90$ and $\omega=10^{-4}$. (a) Oscillating from
lag to anticipatory synchronization via complete synchronization in the region
$t\in(3970,4020)$ and (b) Oscillating from anticipatory to lag
synchronization at the next transition in the region $t\in(27400,27450)$.

\item (a) Difference between $x_1(t-\tau),\tau>0$ and $x_2(t)$,
showing lag synchronization for certain time and (b) difference  between
$x_1(t-\tau), \tau<0$ and  $x_2(t)$, showing anticipatory synchronization for 
the following period of time for  $b_2=0.1$ satisfying the general stability
condition (\ref{eq.four}). Note that complete synchronization occurs in the
transition regime.

\item  The time series $x_1(t-\tau)-x_2(t),\tau<0$, for 
$b_2=0.7$ and $b_3=0.5$ (so that the less stringent condition
$a>|b_2/\sqrt{1-\tau_1^{\prime}}|$ is satisfied while (\ref{eq.four}) is
violated) corresponding to intermittent anticipatory synchronization with the
amplitude of the laminar phase approximately zero.

\item  The time series $x_1(t-\tau)-x_2(t),\tau<0$, for  $b_2=0.1$
and $b_3=1.1$. Here the general stability criterion (\ref{eq.four}) is
satisfied corresponding to intermittent anticipatory synchronization
with the amplitude of the laminar phase exactly zero.

\item The statistical distribution
of laminar phase satisfying $-\frac{3}{2}$ power law scaling for  $b_2=0.1$
and $b_3=1.1$, where the general stability criterion (\ref{eq.four}) is
satisfied.

\item Similarity function for intermittent anticipatory
synchronization $S_a(\tau)$ for different values
of $b_2$, the other system parameters are $a=1.0, b_1=1.2$ and 
$\omega=10^{-5}$. (Curve~1: $b_2=0.1, b_3=1.1$, Curve~2: $b_2=0.7, b_3=0.5$  
 and Curve~3: $b_2=1.1, b_3=0.1$).
 
\item Complete synchronization  between $x_1(t)$ vs $x_2(t)$ when
$\tau_{20}=\tau_{10}$. (a)Approximate complete synchronization for $b_2=0.7$ and
(b) Exact complete synchronization for $b_2=0.1$.

\item  The time series $x_1(t-\tau)-x_2(t),\tau>0$, for  $b_2=0.7$
and $b_3=0.5$ (so that the less stringent condition
$a>|b_2/\sqrt{1-\tau_1^{\prime}}|$ is satisfied while (\ref{eq.four}) is
violated) corresponding to intermittent lag synchronization with the amplitude
of the laminar phase approximately zero. 

\item  The time series $x_1(t-\tau)-x_2(t),\tau>0$, for  $b_2=0.1$
and $b_3=1.1$. Here the general stability criterion (\ref{eq.four}) is
satisfied corresponding to intermittent lag synchronization with the amplitude
of the laminar phase exactly zero.

\item The statistical distribution
of laminar phase satisfying $-\frac{3}{2}$ power law scaling for  $b_2=0.1$
and $b_3=1.1$, where the general stability criterion (\ref{eq.four}) is
satisfied.

\item Similarity function for intermittent lag
synchronization $S_l(\tau)$ for different values
of $b_2$, the other system parameters are $a=1.0, b_1=1.2$ and 
$\omega=10^{-5}$. (Curve~1: $b_2=0.1, b_3=1.1$, Curve~2: $b_2=0.7, b_3=0.5$  
 and Curve~3: $b_2=1.1, b_3=1.0$).

\end{enumerate}
\newpage
\section*{\Large Figures}

\end{document}